\pdfoutput=1
\documentclass[11pt,twoside,a4paper,cmspaper,final,collab]{cms-tdr}
\def\svnVersion{9383cf6}\def\svnDate{2023/07/26}\def\cmsCernNoTag{CERN-EP-2022-178}\def\cmsCernDate{\today}\def\cmsMessage{Published in the European Physical Journal C as \href{http://dx.doi.org/10.1140/epjc/s10052-023-11833-z}{\doi{10.1140/epjc/s10052-023-11833-z}.}}
\begin{document}\cmsNoteHeader{SMP-21-003}

\newlength\cmsFigWidth
\ifthenelse{\boolean{cms@external}}{\setlength\cmsFigWidth{0.85\columnwidth}}{\setlength\cmsFigWidth{0.4\textwidth}}

\newlength\cmsTabSkip\setlength{\cmsTabSkip}{1ex}
\providecommand{\cmsTable}[1]{\resizebox{\textwidth}{!}{#1}}

\newcommand{\PYTHIAE}{{\PYTHIA{8}}\xspace}
\newcommand{\PYTHIAS}{{\PYTHIA{6}}\xspace}
\newcommand{\MG}{\textsc{MG5\_aMC}\xspace}
\newcommand{\MGPyLO}{\textsc{MG5\_aMC+Py8} ($\leq 4j $ LO)\xspace}
\newcommand{\MGPyNLO}{\textsc{MG5\_aMC+Py8} ($\leq 2j $ NLO)\xspace}
\newcommand{\MGTMD}{\textsc{MG5\_aMC+CA3}\xspace}
\newcommand{\MGTMDZj}{\textsc{MG5\_aMC+CA3} (\PZ{+}1 NLO)\xspace}
\newcommand{\MGTMDZjj}{\textsc{MG5\_aMC+CA3} (\PZ{+}2 NLO)\xspace}
\newcommand{\MGTMDMLM}{\textsc{MG5\_aMC+CA3} (\PZ{$\leq 3j$} LO)\xspace}
\newcommand{\cascade}{\textsc{Cascade3}\xspace}
\newcommand{\PBM}{PB\xspace}

\newcommand{\geneva}{\textsc{Geneva} (\PZ{+}0 NNLO)\xspace}

\newcommand{\ptZ}{\ensuremath{\pt(\PZ)}\xspace}
\newcommand{\dphiOneTwo}{\ensuremath{\Delta\phi (j_1j_2)}\xspace}
\newcommand{\dphiZj}{\ensuremath{\Delta\phi (\PZ j_1)}\xspace}

\cmsNoteHeader{SMP-21-003}
\title{Azimuthal correlations in \texorpdfstring{\PZ{}+}{Z+}jets events in proton-proton collisions at \texorpdfstring{$\sqrt{s} = 13\TeV$}{sqrt(s) = 13 TeV}}

\date{\today}

\abstract{ The production of \PZ bosons associated with jets is measured in $\Pp\Pp$ collisions at $\sqrt{s}=13\TeV$ with data recorded with the CMS experiment at the LHC corresponding to an integrated luminosity of 36.3\fbinv. The multiplicity of jets with transverse momentum $\pt > 30\GeV$ is measured for different regions of the \PZ boson's \ptZ, from lower than 10\GeV to higher than 100\GeV. The azimuthal correlation $\Delta \phi$ between the \PZ boson and the leading jet, as well as the correlations between the two leading jets are measured in three regions of \ptZ. The measurements are compared with several predictions at leading and next-to-leading orders, interfaced with parton showers. Predictions based on transverse-momentum dependent parton distributions and corresponding parton showers give a good description of the measurement in the regions where multiple parton interactions and higher jet multiplicities are not important. The effects of multiple parton interactions are shown to be important to correctly describe the measured spectra in the low \ptZ regions.
}
\hypersetup{
pdfauthor={CMS Collaboration},
pdftitle={Azimuthal correlations in Z+jets events in proton-proton collisions at sqrt(s) = 13 TeV},
pdfsubject={CMS},
pdfkeywords={CMS, QCD, anti-kt}}

\maketitle

\section{Introduction}
{\tolerance=1000
In high-energy proton-proton ($\Pp\Pp$) collisions at the CERN LHC, the production of \PZ bosons is regarded as a standard measurement tool, because their properties can be measured very precisely in their leptonic decay channel, and the production cross section can be calculated with high precision. Although the production of \PZ bosons is a purely electroweak (EW) process, corrections from quantum chromodynamics (QCD) play an increasingly important role as the \PZ boson transverse momentum \ptZ  increases. At small \ptZ, where soft-gluon radiation is important, a resummation to all orders must be performed in order to obtain stable theoretical predictions~\cite{Ellis:1981hk,Collins:1984kg,Bizon:2019zgf,Martinez:2019mwt} and to describe the measurements~\cite{Sirunyan:2019bzr}. 
When \ptZ increases, hard partonic radiation becomes important and associated jets can be measured, allowing the study of QCD contributions to \PZ production.
\par}
    
 Cross sections for the production of \PZ bosons associated with jets were measured in proton-antiproton collisions at $\sqrt{s}=1.96\TeV$ at the Fermilab Tevatron by the CDF and \DZERO Collaborations~\cite{Aaltonen:2007ae, D0:2008lbc}.
 At the LHC, the  ATLAS, CMS, and LHCb Collaborations have published measurements in $\Pp\Pp$ collisions at $\sqrt{s}=7\TeV$~\cite{Aad:2013ysa,Aad:2011qv,Khachatryan:2014zya, CMS:2011jak, CMS:2013lua, LHCb:2013mpk}, 8\TeV~\cite{Khachatryan:2016crw,ATLAS:2019bsa}, and 13\TeV~\cite{Sirunyan:2018cpw, ATLAS:2017sag}.

 This article describes a study by the CMS Collaboration of the production of \PZ bosons with associated jets at a center-of-mass energy of 13\TeV.
We measure the multiplicity of jets with $\pt > 30\GeV$ in a pseudorapidity range of $\abs{\eta}<2.4$. 
In the region of low \ptZ, additional jets must balance the leading jet of $\pt > 30\GeV$, whereas at large \ptZ the \PZ boson is expected to balance the \pt of the leading jet. 
   We measure distributions in three (representative) \ptZ regions: at low transverse momentum $\ptZ < 10\GeV$; in the intermediate range  of  $30< \ptZ < 50\GeV$; and in the large range of $\ptZ > 100\GeV$.

{\tolerance=3000
 The  jet multiplicity, the azimuthal correlation \dphiZj between the \PZ boson and the leading jet,  as well as the correlation \dphiOneTwo between the two leading jets, is measured in these three ranges of \ptZ. 
 At small \ptZ, a weak correlation between the \PZ boson and the leading jet is expected, whereas at large \ptZ the azimuthal correlation is expected to be strong, since then the \PZ boson and the leading jet are most likely the highest $\pt$ objects in the event. The situation is opposite for \dphiOneTwo, where at small \ptZ a strong correlation is expected, whereas at large \ptZ the correlation will be weak.

The measurement of jet multiplicity as well as the measurements of the azimuthal correlations \dphiZj, and \dphiOneTwo in various ranges of  \ptZ provide an opportunity to make detailed comparisons with theoretical predictions. In particular, calculations of next-to-leading order (NLO)   \PZ+jet production  supplemented with parton shower (PS) and hadronization, as well as merged calculations with higher partonic jet multiplicity, can be studied.
Of particular interest are the comparisons with predictions based on the
parton branching (\PBM) method with transverse-momentum dependent (\PBM-TMD) parton distribution functions (PDFs)~\cite{Hautmann:2017fcj,Hautmann:2017xtx,Martinez:2018jxt} together with a TMD-based PS~\cite{Baranov:2021uol}. 
 A comparison with resummed calculations using the 
 \textsc{Geneva}~\cite{Alioli:2016wqt,Alioli:2015toa,Alioli:2013hqa,Alioli:2012fc} framework is also shown.
\par}

\section{The CMS detector} \label{sec:detector}

The central feature of the CMS apparatus is a superconducting solenoid of 6\unit{m} internal diameter, providing a magnetic field of 3.8\unit{T}. Within the solenoid volume are a silicon pixel and strip tracker, a lead tungstate crystal electromagnetic calorimeter (ECAL), and a brass and scintillator hadron calorimeter (HCAL), each composed of a barrel and two endcap sections. Forward calorimeters extend the $\eta$ coverage provided by the barrel and endcap detectors. Muons are detected in gas-ionization chambers embedded in the steel flux-return yoke outside the solenoid. 

Events of interest are selected using a two-tiered trigger system. The first level, composed of custom hardware processors, uses information from the calorimeters and muon detectors to select events at a rate of around 100\unit{kHz} within a fixed latency of about 4\mus~\cite{Sirunyan:2020zal}. The second level, known as the high-level trigger, consists of a farm of processors running a version of the full event reconstruction software optimized for fast processing and reduces the event rate to around 1\unit{kHz} before data storage~\cite{Khachatryan:2016bia}.

The particle-flow algorithm (PF)~\cite{CMS-PRF-14-001} reconstructs and identifies each individual particle in an event, with an optimized combination of information from the various elements of the CMS detector.
The primary vertex (PV) is taken to be the vertex corresponding to the hardest scattering in the event, evaluated using tracking information alone, as described in Section 9.4.1 of Ref.~\cite{CMS-TDR-15-02}.

The energy of photons is obtained from the ECAL measurement. The energy of electrons is determined from a combination of the electron momentum at the primary interaction vertex as determined by the tracker, the energy of the corresponding ECAL cluster, and the energy sum of all bremsstrahlung photons spatially compatible with originating from the electron track. The momentum resolution for electrons with $\pt \approx 45\GeV$ from $\PZ \to \Pep \Pem$ decays ranges from 1.7--4.5\%. It is generally better in the barrel region than in the endcaps, and also depends on the bremsstrahlung energy emitted by the electron as it traverses the material in front of the ECAL~\cite{CMS-DP-2020-021,Sirunyan_2021}. The overall reconstruction efficiency is around 93\% for electrons from \PZ decay.

The energy of muons is obtained from the curvature of the corresponding track. Muons are measured in the range $\abs{\eta}<2.4$, with detection planes made using three technologies: drift tubes, cathode strip chambers, and resistive-plate chambers. The single-muon trigger efficiency exceeds 90\% over the full $\eta$ range, and the efficiency to reconstruct and identify muons is greater than 96\%. Matching muons to tracks measured in the silicon tracker results in a relative \pt resolution of 1\% in the barrel and 3\% in the endcaps for muons with \pt up to 100\GeV. The \pt resolution in the barrel is better than 7\% for muons with \pt up to 1\TeV~\cite{Sirunyan:2018fpa}. 

The energy of charged hadrons is determined from a combination of their momentum measured in the tracker and the matching ECAL and HCAL energy deposits, corrected for the response function of the calorimeters to hadronic showers. Finally, the energy of neutral hadrons is obtained from the corresponding corrected ECAL and HCAL energies.

{\tolerance=1000
  For each event, hadronic jets are clustered from these reconstructed particles using the infrared and collinear safe anti-\kt algorithm~\cite{Cacciari:2008gp, Cacciari:2011ma} with a distance parameter of 0.4.
Jet momentum is determined as the vectorial sum of all particle momenta in the jet and is found from simulation to be, on average, within 5 to 10\% of the true momentum over the whole \pt spectrum and detector acceptance. Additional $\Pp\Pp$ interactions within the same or nearby bunch crossings (pileup) contribute additional tracks and calorimetric energy depositions, increasing the apparent jet momentum. To mitigate this effect, tracks identified as originating from pileup vertices are discarded and a correction is applied to correct for any remaining contributions. 

Jet energy corrections are derived from simulation studies so the average measured energy of jets becomes identical to that of particle level jets. In situ measurements of the momentum balance in dijet, photon+jet, \PZ{}+jet, and multijet events are used to determine any residual differences between the jet energy scale in data and in simulation, and appropriate corrections are made~\cite{Khachatryan:2016kdb}. Additional selection criteria are applied to each jet to remove jets potentially dominated by instrumental effects or reconstruction failures. 
The jet energy resolution amounts typically to 15--20\% at 30\GeV, 10\% at 100\GeV, and 5\% at 1\TeV~\cite{Khachatryan:2016kdb}. 
\par}

During the 2016 data-taking, a gradual shift in the timing of the inputs of the ECAL first-level trigger in the region at $\abs{\eta} > 2.0$, referred to as prefiring, caused a specific trigger inefficiency. For events containing an electron (a jet) with \pt larger than 50 (100)\GeV, in the region $2.5 < \abs{\eta} < 3.0$ the efficiency loss is about 10--20\%, depending on \pt, $\eta$, and timing. Correction factors were computed from data and applied to the acceptance evaluated by simulation.

{\tolerance=1000
A more detailed description of the CMS detector, together with a definition of the coordinate system used and the relevant kinematic variables, is reported in Ref.~\cite{Chatrchyan:2008zzk}. 
\par}

\section{Theoretical predictions}
\label{sec:theory}
                            
{\tolerance=1000
The measured differential cross sections are compared with a variety of predictions.
    One of the NLO calculations uses \MGvATNLO~\cite{Alwall:2014hca} (version 2.2.2) event generator interfaced with \PYTHIAE~\cite{Sjostrand:2014zea} for PS and hadronization. The matrix element calculations include $\PZ/\PGg^{*}$+0,1,2 jets at NLO.
It is labeled as \MGPyNLO in the following.
\par}

{\tolerance=1000
 The measurements are also compared with predictions obtained from \MGvATNLO (version 2.6.9) with \PBM-TMD PDFs and the corresponding PS as implemented in \cascade~\cite{Baranov:2021uol} (labeled as \MGTMD ).  The matrix elements (MEs) are calculated at NLO for \PZ{}+1 and \PZ{}+2 partons separately.
   The parton density  \textsc{PB-NLO-HERA}\uppercase\expandafter{\romannumeral1}$+$\uppercase\expandafter{\romannumeral2}-\textsc{2018-set2}~\cite{Martinez:2018jxt}, as well as the \PBM  initial-state PS, follow angular ordering conditions~\cite{Dokshitzer:1987nm,Bassetto:1983mvz,Marchesini:1987cf,Catani:1990rr}. 
 The advantage of the  \MGTMD calculation is that the parameters of the  \PBM-TMD initial-state PS are fixed by the \PBM-TMD PDFs.
\par}

In all calculations using \MGvATNLO\ the renormalization and factorization scales are set to $\mu_\mathrm{r}=\mu_\mathrm{f}=1/2 \sum_i \mathrm{H}_{\mathrm{T}\,i}$, where $\mathrm{H}_{\mathrm{T}\,i}$ is the scalar sum of the \pt with $i$ running over all final particles and partons in the ME calculation. The corresponding uncertainties are estimated as the envelope of the set of variations of  $\mu_\mathrm{r}$ and $\mu_\mathrm{f}$ by factors of 2 and 1/2, in all possible combinations except the extreme cases $(\mu_\mathrm{f} , \mu_\mathrm{r})=(2,0.5), (0.5,2)$.
The PDF uncertainties are estimated as the standard deviation of observables when using weights from the replicas provided in the NNPDF 3.0 NLO~\cite{Ball:2014uwa} PDF set.
  
The corresponding versions of these generators at leading-order (LO) are also compared with the measurement.

The following calculations are used for comparison with the measurements (a summary is given in Table~\ref{tab:MC}):

{\tolerance=2000
\begin{itemize}
 
\item \MGPyNLO   is a fixed-order perturbative QCD calculation at NLO of up to 2 noncollinear high-\pt partons  for $\Pp\Pp \to \PZ$+$N$, $N=0$, 1, 2, supplemented with PS and multiparton interactions (MPIs) from \PYTHIAE  (version~8.212). 
The parameters of the underlying event tune CUETP8M1~\cite{Khachatryan:2015pea} are applied.
The merging of PS and MEs is performed
with the FxFx scheme~\cite{Frederix:2012ps} with the merging scale of 30\GeV and a minimal partonic \pt for jets of $\pt^{\text{part}} = 15\GeV$. 
  The NNPDF 3.0 NLO PDFs are used and $\alpS(m_\PZ) =0.118$ is chosen, where $m_\PZ$ is the \PZ boson mass. 
The predictions from \MGPyNLO are used to investigate the effect of MPI.

\item \MGPyLO includes MEs computed at LO for  $\Pp\Pp \to \PZ $+$N$ partons, $N=0,1\ldots4$, using the \kt-MLM~\cite{Alwall_2007} procedure to match the different parton multiplicities of the MEs to the PS, with the matching scale set to 19\GeV. The \PYTHIAE generator (version~8.212) is interfaced with \MG to include initial- and final-state PS and hadronization, with settings defined by the CUETP8M1 tune~\cite{Khachatryan:2015pea}. The \textsc{NNPDF 2.3 LO}~\cite{Ball:2013hta,Ball:2011uy} PDF is used, and the strong coupling $\alpS(m_\PZ )$ is set to 0.130.
The total cross section for $\PZ{}\to \Pell^+ \Pell^- + \geq$0 jet is normalized to the predictions of \FEWZ v3.1 next-to-next-to-leading order (NNLO)~\cite{Li_2012} applying a K-factor of 1.17. 

\item \MGTMDZj is a fixed-order perturbative QCD calculation at NLO of one noncollinear high-\pt parton for $\Pp\Pp \to \PZ $+1 with $\pt^\text{part}>15\GeV$, supplemented with \PBM-TMD PDFs and PS, which for the initial state follows the \PBM-TMD distribution~\cite{Baranov:2021uol}.
 The NLO \PBM-TMD~set~2~\cite{Martinez:2018jxt} with $\alpS(m_\PZ )= 0.118$ is used, the collinear version of  \PBM set~2 is used for the ME calculation. 
This leads to cross sections 10--20\% smaller than obtained with other PDFs  because \PBM parton densities are determined from a fit to HERA data only.
Therefore, an overall normalization factor of 1.2 is applied to the \PBM prediction.
 The inclusion of the transverse momentum \kt and initial-state PS is performed with \textsc{Cascade3}~\cite{Baranov:2021uol} (version 3.2.1). Final-state radiation, which is not constrained by the \PBM-TMD PDF, and hadronization is performed with \PYTHIAS (version 6.428)~\cite{Sjostrand:2006za}. 
  MPI effects are not simulated in this approach.

\item \MGTMDZjj  is a fixed-order perturbative QCD calculation at NLO of two noncollinear high-\pt partons for $\Pp\Pp \to \PZ $+2 with $\pt^\text{part}>15\GeV$, supplemented with \PBM-TMD PDFs and parton showering and hadronization.
The same \PBM-TMD distribution and PS as in \MGTMDZj is applied.

\item \MGTMDMLM uses \textsc{MG5\_aMC} to generate $\PZ $+0,1,2,3 jet samples at LO with a partonic generation cut $\pt^\text{part}> 15\GeV$. The TMD merging~\cite{BERMUDEZMARTINEZ2021136700} procedure for combining the TMD PS with the ME calculations is used.
 The same \PBM-TMD distributions and PS as in \MGTMDZj are applied.
A merging scale value of 23\GeV is used, since it provides a smooth transition between ME and PS computations. An overall  K-factor of 1.27 is applied to the prediction. MPI effects are not simulated.

\item \geneva (1.0-RC3)~\cite{Alioli:2016wqt,Alioli:2015toa,Alioli:2013hqa,Alioli:2012fc} 
is based on NNLO calculations for the processes $\Pp\Pp \to \PZ{}/\gamma \to \Pep \Pem$ and $\PGmp \PGmm$  combined with higher-order resummation. The calculation uses the PDF4LHC15 NNLO PDF set~\cite{Butterworth_2016} with $\alpS(m_\PZ )= 0.118$.
The simulation of PS, hadronization and MPI is performed by \PYTHIAE (version~8.212) with the CUETP8M1 tune.

\end{itemize}
\par}

\begin{table*}
    \centering  
       \topcaption{Description of the simulated samples used in the analysis.}
    \cmsTable{
    \begin{tabular}{l l l l} 
        Generator & PDF & Matrix element & Tune \\
        \hline
       \MGPyNLO~\cite{Alwall:2014hca}  & $\mathrm{NNPDF3.0}$ (NLO)~\cite{Ball:2014uwa}  & NLO ($2\to\PZ{}$+0,1,2) & $\mathrm{CUETP8M1}$~\cite{Khachatryan:2015pea} \\
        \MGPyLO~\cite{Alwall:2011uj,Alwall:2014hca}  & $\mathrm{NNPDF2.3}$ (LO)~\cite{Ball:2011uy} & LO ($2\to\PZ$+0,1,2,3,4) & $\mathrm{CUETP8M1}$~\cite{Khachatryan:2015pea} \\
       \MGTMDZj~\cite{Alwall:2014hca}  & $\mathrm{\PBM~NLO~set2}$ (NLO)~\cite{Martinez:2018jxt}& NLO ($2\to\PZ$+1) & \NA \\
       \MGTMDZjj~\cite{Alwall:2014hca}  & $\mathrm{\PBM~NLO~set2}$ (NLO)~\cite{Martinez:2018jxt}& NLO ($2\to\PZ$+2) & \NA \\
       \MGTMDMLM~\cite{Alwall:2014hca,BERMUDEZMARTINEZ2021136700}  & $\mathrm{PB~NLO~set2}$ (NLO)~\cite{Martinez:2018jxt}& LO ($2\to\PZ$+0,1,2,3) & \NA \\
       \geneva~\cite{Alioli:2016wqt,Alioli:2015toa,Alioli:2013hqa,Alioli:2012fc} &  $\mathrm{NNPDF3.1}$ (NLO)~\cite{Ball:2017nwa} & NNLO ($2\to\PZ$) & $\mathrm{CUETP8M1}$~\cite{Khachatryan:2015pea} \\
      \end{tabular}
    }
    \label{tab:MC}
\end{table*}

\subsection{Simulated samples}
\label{sec:samples}
Events generated by \MGPyNLO  
 are passed through a full detector simulation based on \GEANTfour~\cite{geant}. The simulated events are reconstructed using standard CMS reconstruction packages. This sample is used for the simulation of the signal process to estimate efficiencies, systematic uncertainties 
and for the correction of the data for detector spreading effects and inefficiencies, the so-called unfolding procedure.
 
{\tolerance=1000
Other processes that can give a final state with two oppositely charged same-flavor leptons and jets are
\ttbar, single top, vector boson pair (VV) and W+jets.
The \ttbar and single top backgrounds are generated using POWHEG 2.0 
 ~\cite{Nason:2004rx,Frixione:2007vw,Alioli:2010xd, Re:2010bp, Alioli:2009je, Frixione:2007nw} 
 interfaced with \PYTHIAE.
The total cross section of \ttbar production is normalized to the prediction with \textsc{NNLO} accuracy in QCD and next-to-next-to-leading logarithmic (NNLL) gluon radiation resummation calculated with \textsc{Top$++$~2.0}~\cite{Czakon:2011xx}.
The double vector boson productions are generated with \MG~($\PW\PZ$),  \textsc{Powheg}~($\PW\PW$), both interfaced to \PYTHIAE, or with \PYTHIAE for $\PZ\PZ$.
The total cross sections for the $\PW\PZ$ and $\PZ\PZ$ diboson samples are normalized to the NLO prediction calculated with \textsc{MCFM~6.6}~\cite{2010}.
The W+jets sample is generated by \MG at NLO accuracy, interfaced with \PYTHIAE.
The \PZ boson decay into $\PGt^{+}\PGt^{-}$ is included in the signal simulation and considered as a background.
\par}

\section{Data analysis} \label{sec:analysis}

The differential cross section of \PZ bosons with associated jets is measured in bins of \ptZ, as functions of the jet multiplicity, the azimuthal angles  \dphiZj and \dphiOneTwo, where \ensuremath{j_1} is the leading jet and \ensuremath{j_2} is the second-leading jet.
 
\subsection{Event selection}\label{sec:selection}

The data samples recorded in 2016 correspond to an integrated luminosity of 36.3\fbinv.
  Events with a pair of leptons ($\PGmp\PGmm$ or $\Pep\Pem$) consistent with the decay of a \PZ boson and with jets reconstructed from PF candidates are selected as \PZ{}+jet events.
  Those events are required to pass a series of selection criteria to reduce the background contributions.
An event is selected if the double muon (electron) trigger
 with 18 and 7 (23 and 12)\GeV thresholds in \pt or a single muon trigger with a threshold of 24\GeV is satisfied.
In the offline selection, the leading (subleading) electron and muon candidates must have transverse momenta of $\pt > 25\,(20)\GeV$ in a range of $\abs{\eta} < 2.4$. Only events with pairs of  oppositely charged muons (electrons) with an invariant mass in the range $91\pm15\GeV$ are accepted.

{\tolerance=4800
 Muon candidates are required to be isolated from other particles, as specified by an isolation criteria, $I_\mathrm{ISO}$: 
\begin{linenomath}
\ifthenelse{\boolean{cms@external}}
{ 
\begin{multline*}
    I_\mathrm{ISO}=\bigg[\sum^\text{charged} \pt + \max\Big(0, \sum^\text{neutral} \pt + \sum^\mathrm{EM} \pt - \\ 
0.5 \sum^\mathrm{PU} \pt\Big)\bigg]/\pt^{\PGm} \leq 0.15,
\end{multline*}
} 
{ 
\begin{equation*}
    I_\mathrm{ISO}=\bigg[\sum^\text{charged} \pt + \max\Big(0, \sum^\text{neutral} \pt + \sum^\mathrm{EM} \pt - 0.5 \sum^\mathrm{PU} \pt\Big)\bigg]/\pt^{\PGm} \leq 0.15,
\end{equation*}
} 
\end{linenomath}
where the sums run over the corresponding particles inside a cone of radius $\Delta R=\sqrt{\smash[b]{(\Delta\eta)^2+(\Delta\phi)^2}}=0.4$  
around the muon candidate considering separately  charged hadrons ($\text{charged}$), neutral hadrons ($\text{neutral}$), photons ($\mathrm{EM}$), and charged particles from pileup ($\mathrm{PU}$). 
\par}
 
Electrons are required to be isolated from other particles, as specified by an isolation criteria, $I_\mathrm{ISO}$: 
\begin{linenomath}
\ifthenelse{\boolean{cms@external}}
{ 
\begin{multline*}
    I_\mathrm{ISO}=\bigg[\sum^\text{charged} \pt + \max\Big(0, \sum^\text{neutral} \pt + \sum^\mathrm{EM} \pt -\\
 \rho \mathrm{A}_\text{eff}\Big)\bigg]/\pt^{\Pe}  \leq 0.15,   
\end{multline*}
} 
{ 
\begin{equation*}
    I_\mathrm{ISO}=\bigg[\sum^\text{charged} \pt + \max\Big(0, \sum^\text{neutral} \pt + \sum^\mathrm{EM} \pt - \rho \mathrm{A}_\text{eff}\Big)\bigg]/\pt^{\Pe}  \leq 0.15,   
\end{equation*}
} 
\end{linenomath}
\begin{equation}  
\end{equation} 
where the sums run over the corresponding particles inside a cone of radius $\Delta R = 0.3$.
The term $\rho \mathrm{A}_\text{eff}$ represents a correction for pileup effects, where $\rho$ corresponds to the amount of \pt added to the event per unit area and $\mathrm{A}_\text{eff}$ is the area of the isolation region weighted by a factor that accounts for the dependence of the pileup transverse energy density on the electron $\eta$~\cite{2015,Sirunyan_2021}.
 
 Jets are required to have a minimum \pt of 30\GeV to ensure that they are well measured and to reduce the pileup contamination. Jets are limited to a rapidity range of $\abs{y}<2.4$, and are required to be isolated from the lepton candidates by $\Delta R_{\Pell,j}>0.4$.
To keep only charged particles originating from the \PZ boson vertex, charged particles identified as originating from pileup vertices are discarded. 
As discussed in Section~\ref{sec:detector}, jet energy corrections are applied to data and simulation.
The jet energy resolution (JER) in simulation is further spread to match that in data. 

 The simulated events are reweighted such that their pileup distribution matches the measured one in each data-taking period.

\begin{figure*}[htbp!]
 \centering
    \includegraphics[width=0.38\textwidth]{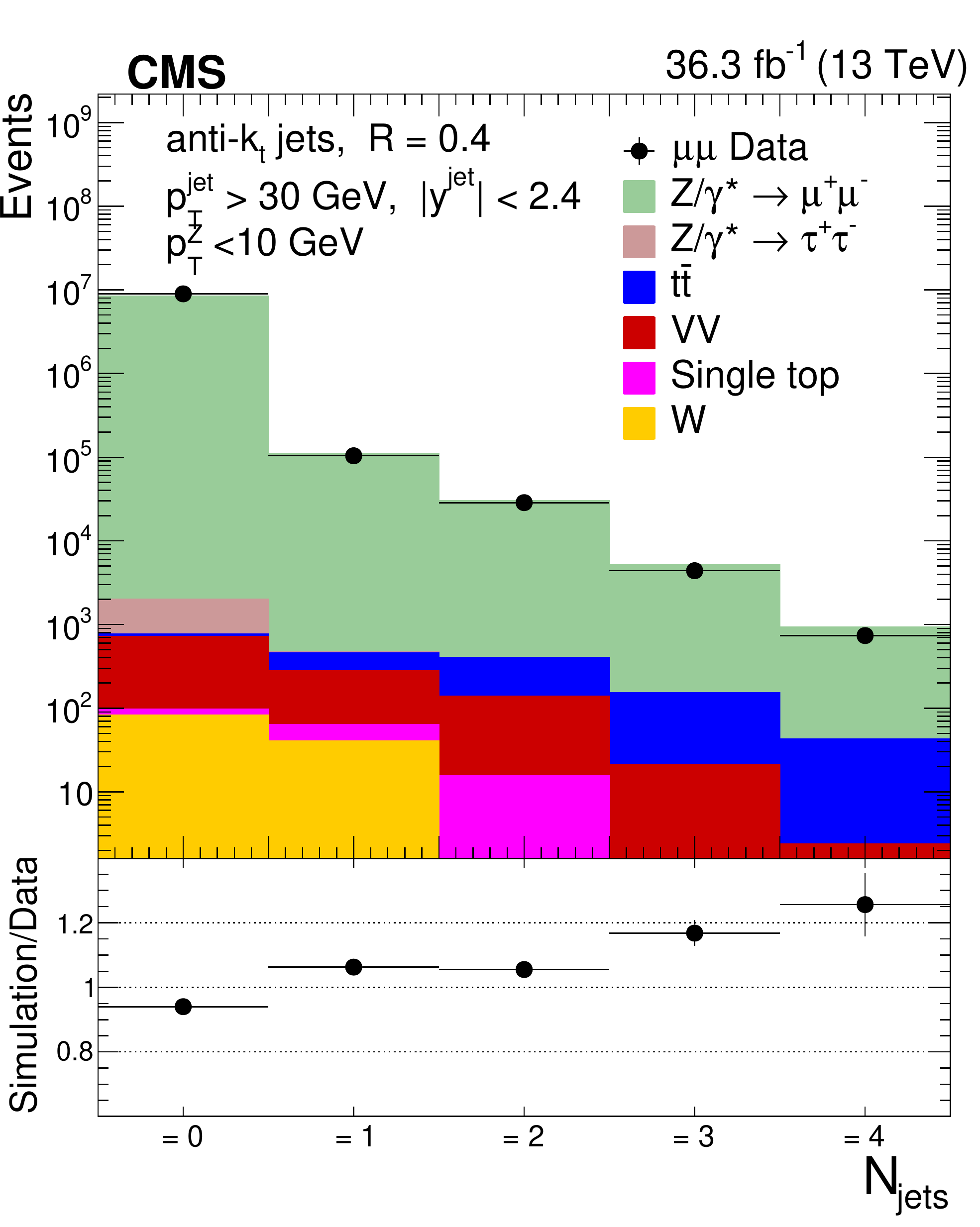}
    \includegraphics[width=0.38\textwidth]{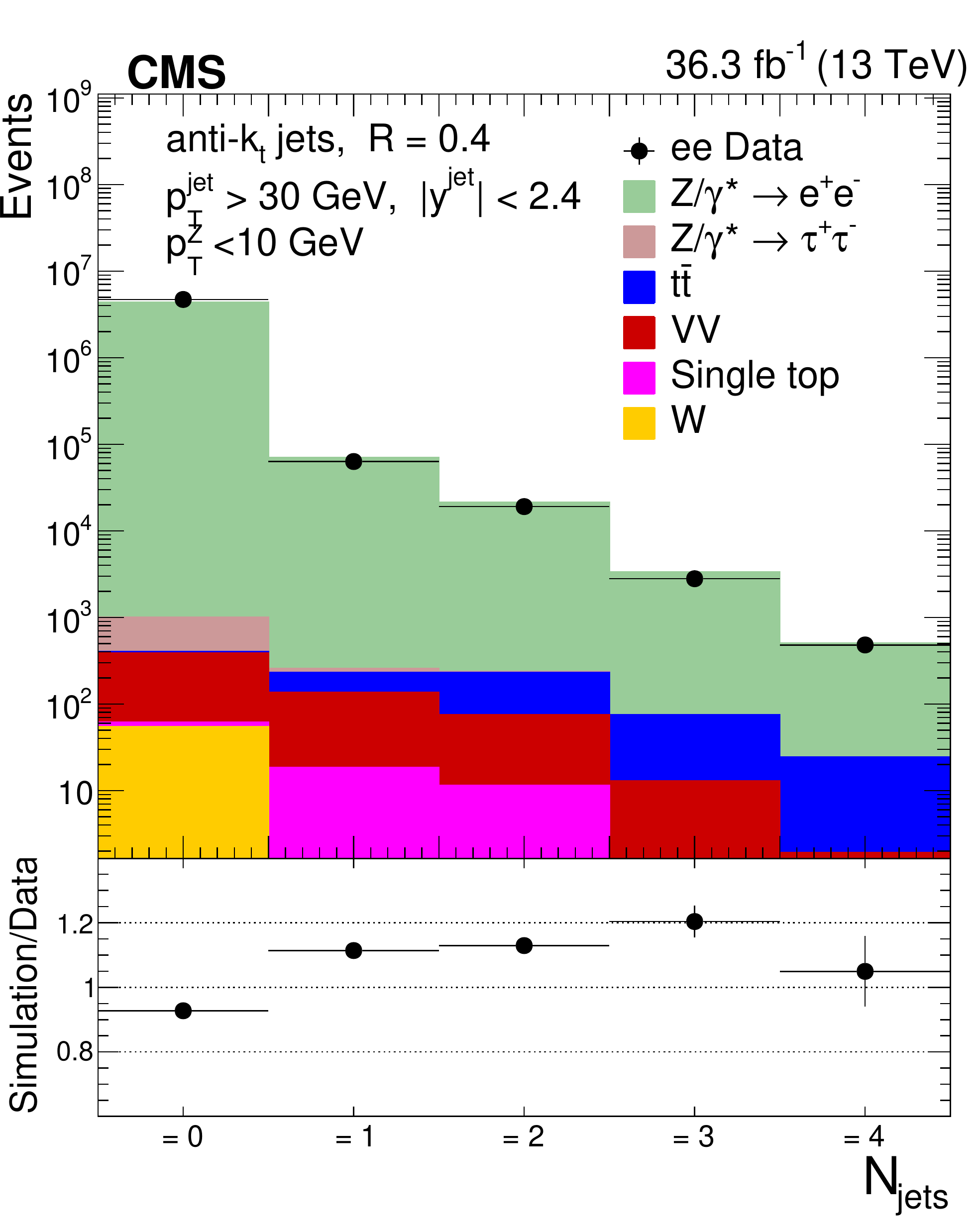}
    \includegraphics[width=0.38\textwidth]{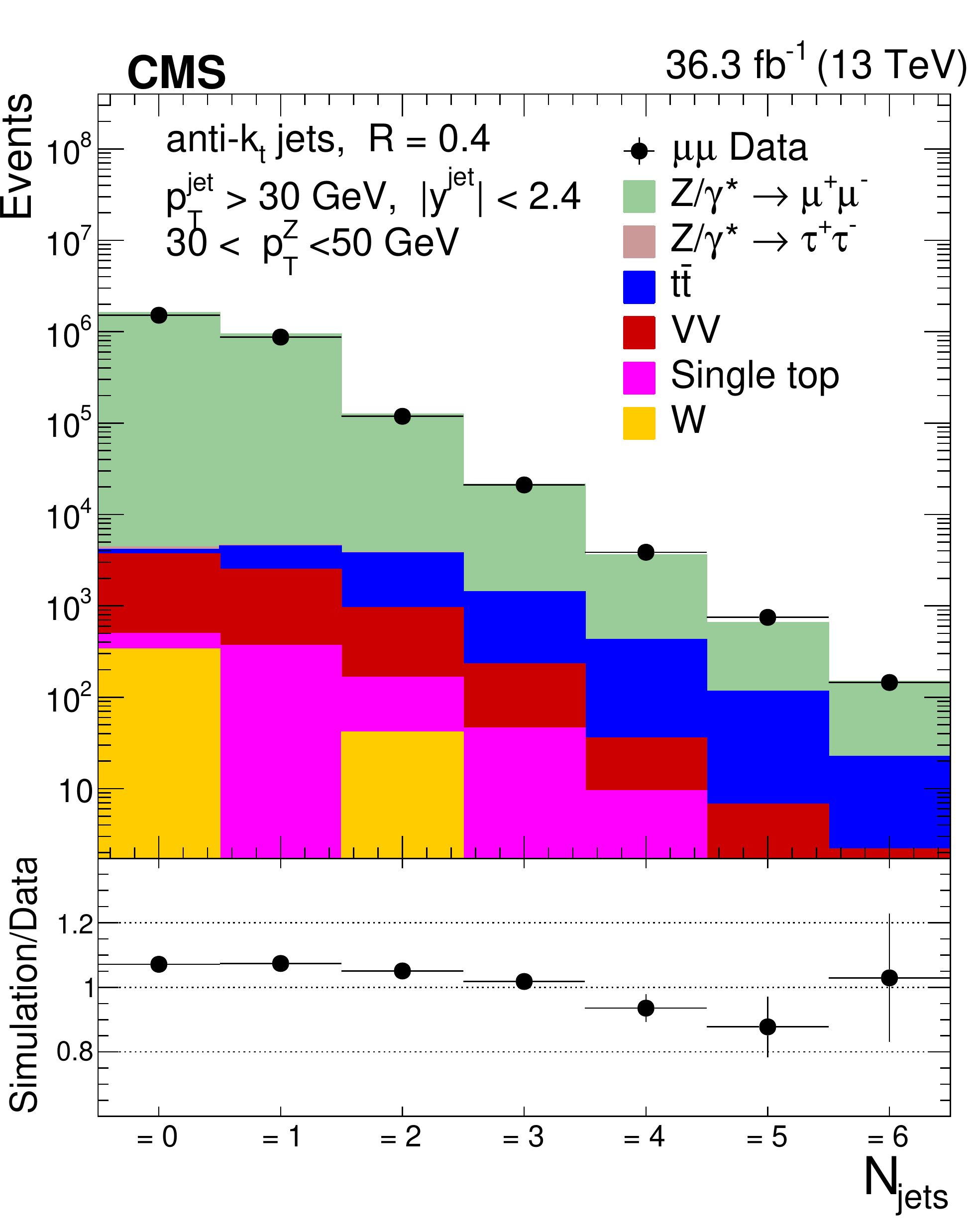}
    \includegraphics[width=0.38\textwidth]{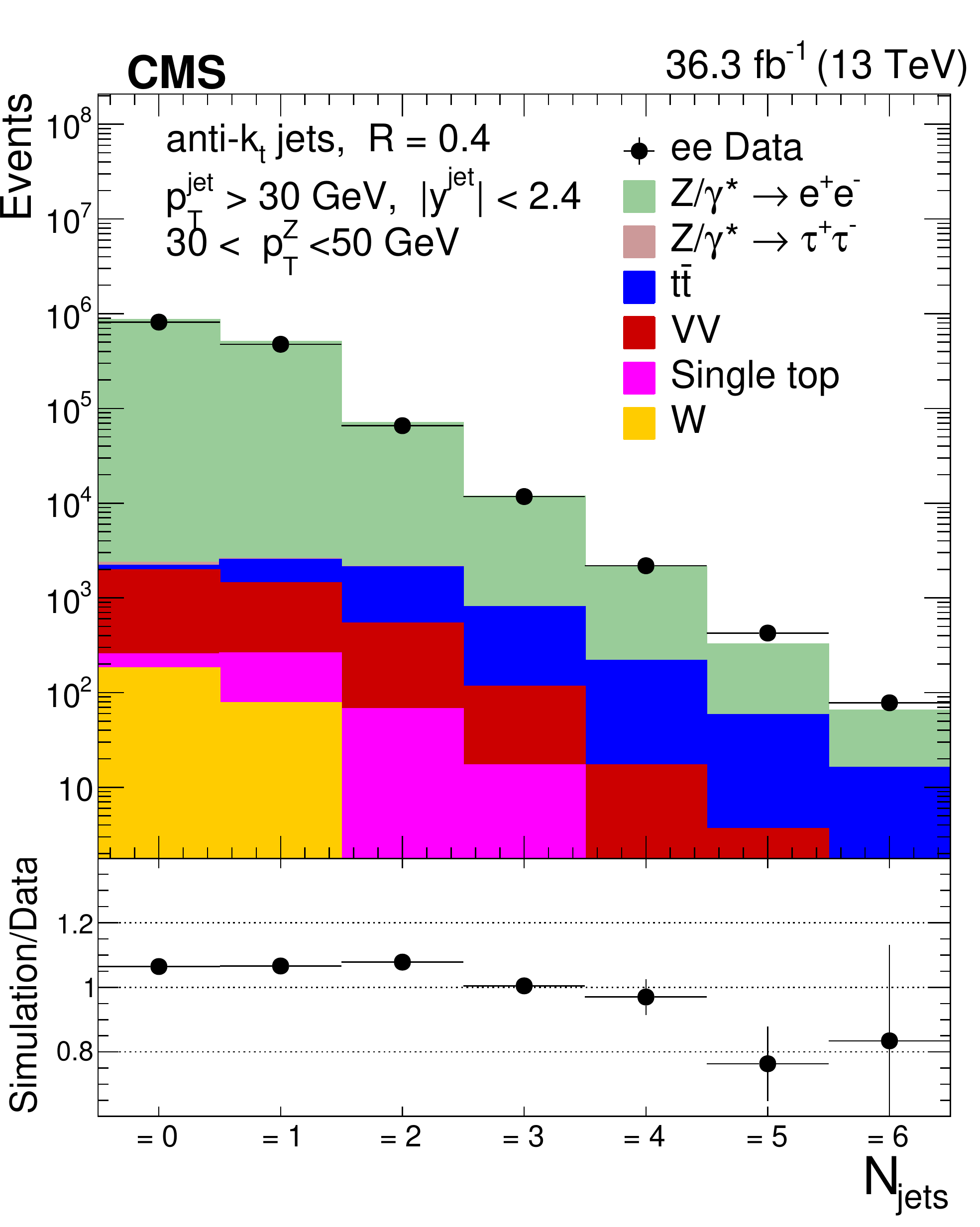}
    \includegraphics[width=0.38\textwidth]{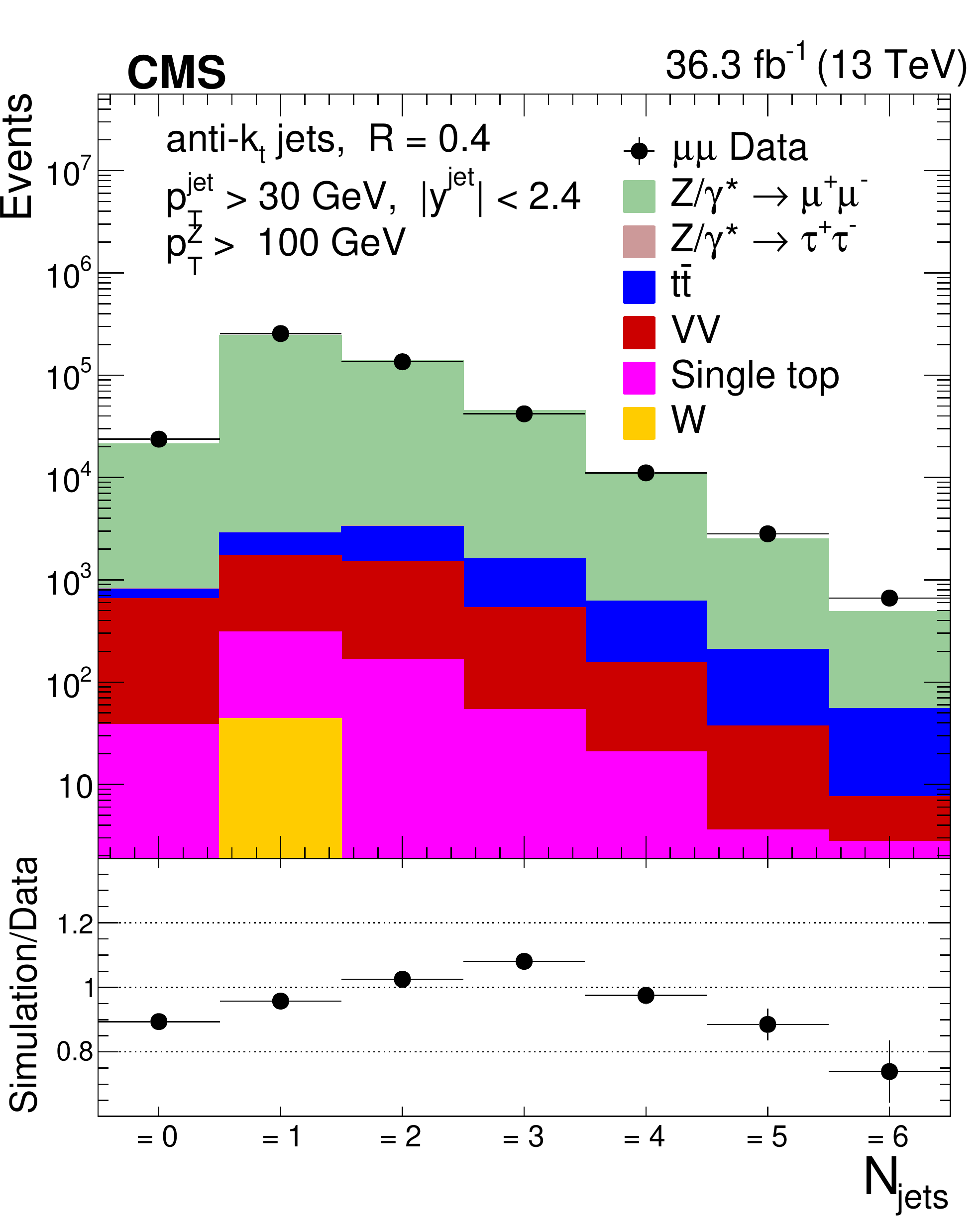}
    \includegraphics[width=0.38\textwidth]{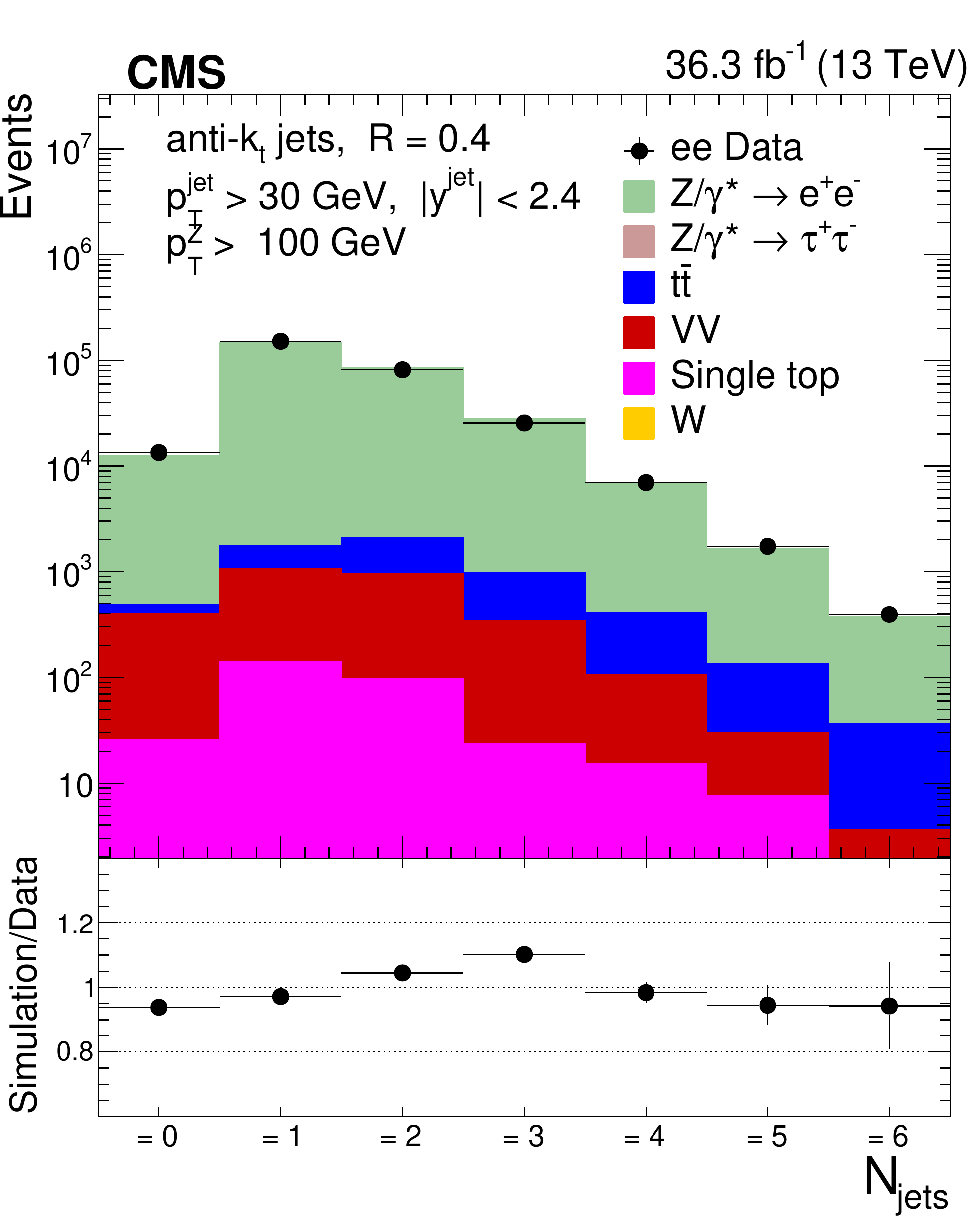}
    \topcaption{
     The exclusive jet multiplicity distribution before unfolding in three different regions of \protect\ptZ : 
$\ptZ<10\GeV$ (upper), $30<\ptZ<50\GeV$ (middle),   $\ptZ>100\GeV$ (lower) for the \MM channel  (left) and the \EE channel (right). The error bars around the data points represent the statistical uncertainties. }
     \label{fig:Zmass}
\end{figure*}

 Several corrections for leptons are applied to the simulation yields to compensate for the measured differences between the efficiencies in data and simulation.
These corrections are applied as
 trigger, lepton identification, and lepton isolation scale factors. 
The values of the scale factors are close to one.
An additional trigger inefficiency correction due to the prefiring effect is included.
   The exclusive jet multiplicity in different regions of \protect\ptZ for muon and electron channels is shown in Fig.~\ref{fig:Zmass}.

\subsection{Correction for the detector effects}\label{sec:correction}

 Detector effects, like inefficiencies and the spreading of the particle momentum, energy and angle, are corrected using the an unfolding procedure, which is applied after background subtraction. 
  The iterative D'Agostini method as implemented in RooUnfold~\cite{DAgostini:1994fjx,RooUnfold} is used. 
 The iteration is affected by fluctuations that increase with the number of iterations. The fluctuations are studied for each distribution and the procedure of unfolding is stopped before the fluctuations become significant with respect to the statistical uncertainty, following the method used in~\cite{Sirunyan:2018cpw}.
Through the unfolding procedure the cross section at the stable-particle level is obtained.
  Particles are considered stable if their proper lifetime is above 10\mm/c.
Neutrinos are not included.
The momentum of the leptons is calculated including photons in a cone of a radius of $\Delta R=0.1$ (``dressed" leptons).
The phase space definition for the final cross sections is given in Table~\ref{tab:PhaseSpaceHad}.

\begin{table}[htb]
    \centering
    \topcaption{Particle-level phase space definition}
    \label{tab:PhaseSpaceHad}
     \begin{tabular}{l l l } 
        object &{requirement} \\
        \hline
        leading (subleading) lepton  &  $\pt>25(20)\GeV$, $\abs{\eta}<2.4$ \\
        lepton-jet separation & $\Delta R_{\Pell,j}>0.4 $ \\
        lepton pair mass &  {$76<m_{\Pell^+ \Pell^-}<106\GeV$}\\
        jet          & $\pt>30\GeV$,  $\abs{\eta_{jet}}<2.4$\\
    \end{tabular}
\end{table}

\subsection{Background estimation}

       The contributions from background processes are estimated using MC-based simulations, described in Section~\ref{sec:samples}, and are subtracted from the measured distributions.
The dominant background, $\PQt\PAQt$, is verified with data control samples, using the same criteria as for the measurement, but requiring the two leptons to have different flavours ($\Pe \PGm$ instead of $\PGmp \PGmm$ or $\Pep \Pem$).
The effect of mismodeling of top quark distributions is covered by the MC uncertainties. Therefore, no additional correction or uncertainty is applied~\cite{Khachatryan:2016crw}.
 The $\PZ{}\to \PGt^{+}\PGt^{-}$ decays are considered as a background, and their contribution is estimated from simulation and subtracted during the unfolding procedure.

\subsection{Uncertainties}\label{sec:uncertainties}

 The statistical uncertainties from the measured observables are propagated to the final results via the unfolding procedure.
The systematic uncertainties originate from the following sources: 
 \begin{itemize}
 \item Jet energy scale:\\
Variations of the jet energy scale corrections~\cite{Khachatryan:2016kdb,CMS-PAS-JME-16-003} are applied as functions of  \pt and $\eta$ for individual run periods; this affects the differential cross sections by 3--7\%.
 \item Jet energy resolution:\\
The JER~\cite{Khachatryan:2016kdb,CMS-PAS-JME-16-003} uncertainty is obtained by varying the spreading factor to match the simulated jet energy resolution to data by one standard deviation around its central value, resulting in an uncertainty of up to 1--2\%.
  \item Efficiency correction:\\
The uncertainty coming from the measurements of trigger efficiency, lepton reconstruction, and lepton identification is estimated by varying the scale factors by their uncertainties, as described in Ref.~\cite{Sirunyan:2019bzr}. 
The resulting uncertainty in the differential cross section measurement is less than 1\%.
 \item Luminosity:\\
{\tolerance=1000
The uncertainty in the integrated luminosity
is 1.2\%~\cite{CMS:2021xjt}.
 It is applied as a global scale factor to the cross section as well as to the normalization of the background samples.
\par}
 \item Pileup:\\
The determination of the simulated pileup profile is based on a total inelastic pp cross section of 69.2 mb~\cite{2018PU}.
Alternative pileup profiles are generated by varying this cross section by 5\% affecting the measurement by 1--2\%.
   \item Prefiring:\\
The prefiring uncertainty is estimated by up and down variations of the prefiring weight. The uncertainty is less than 0.5\%.
\item Background:\\
    The theoretical uncertainty in the calculation of background processes is used to estimate the uncertainty in the background modeling.
The main source of background is  $\PQt\PAQt$ production on this process.  An uncertainty of  around 6\% is estimated 
using the \textsc{TOP++2.0} program, which includes scale and PDF variations. The resulting uncertainty is less than 0.2\%.
Systematic uncertainties stemming from other background processes are negligible.
 \item Unfolding and model:\\
The uncertainty of unfolding and modeling
is estimated by reweighting the simulated signal event sample to match the data and using this as an alternative model for unfolding.
This gives an uncertainty of about 2\%. 
The uncertainty coming from the finite size of the simulation sample that is used 
to correct the data for detector effects
 results in an uncertainty of 2--8\%.
\end{itemize}
A summary table of these uncertainties is given in Table~\ref{tab:unc}.
All the systematic uncertainties are quadratically summed assuming independent uncertainty sources.

 \begin{table}
    \topcaption{Systematic uncertainties on the unfolded differential cross section}
        \centering
    \begin{tabular}{l l}
    Uncertainty source & [\%]\\
        \hline
    Jet energy scale & 3--7\\
    Jet energy resolution & 1--2\\
    Efficiency correction & $<$1 \\
    Luminosity & 1.2 \\
    Pileup & 1--2\\
    Prefiring & $<$0.5 \\
    Background & $<$0.2 \\
    Unfolding and model & 2--8 \\ [\cmsTabSkip]
    Total & 4--11 \\
    \end{tabular}
           \label{tab:unc}
\end{table}

\section{Results} \label{sec:results}
The production cross section of \PZ{}+jets is measured in the phase space given in Table~\ref{tab:PhaseSpaceHad}. 
The \PZ boson is identified via its leptonic decay channel. 
 The results of the muon and electron decay channels are combined using the best linear unbiased estimator (BLUE)~\cite{Lyons:1988rp,Valassi:2003mu} approach.

\begin{figure*}[htp]
\centering
  \includegraphics[width=0.488\textwidth]{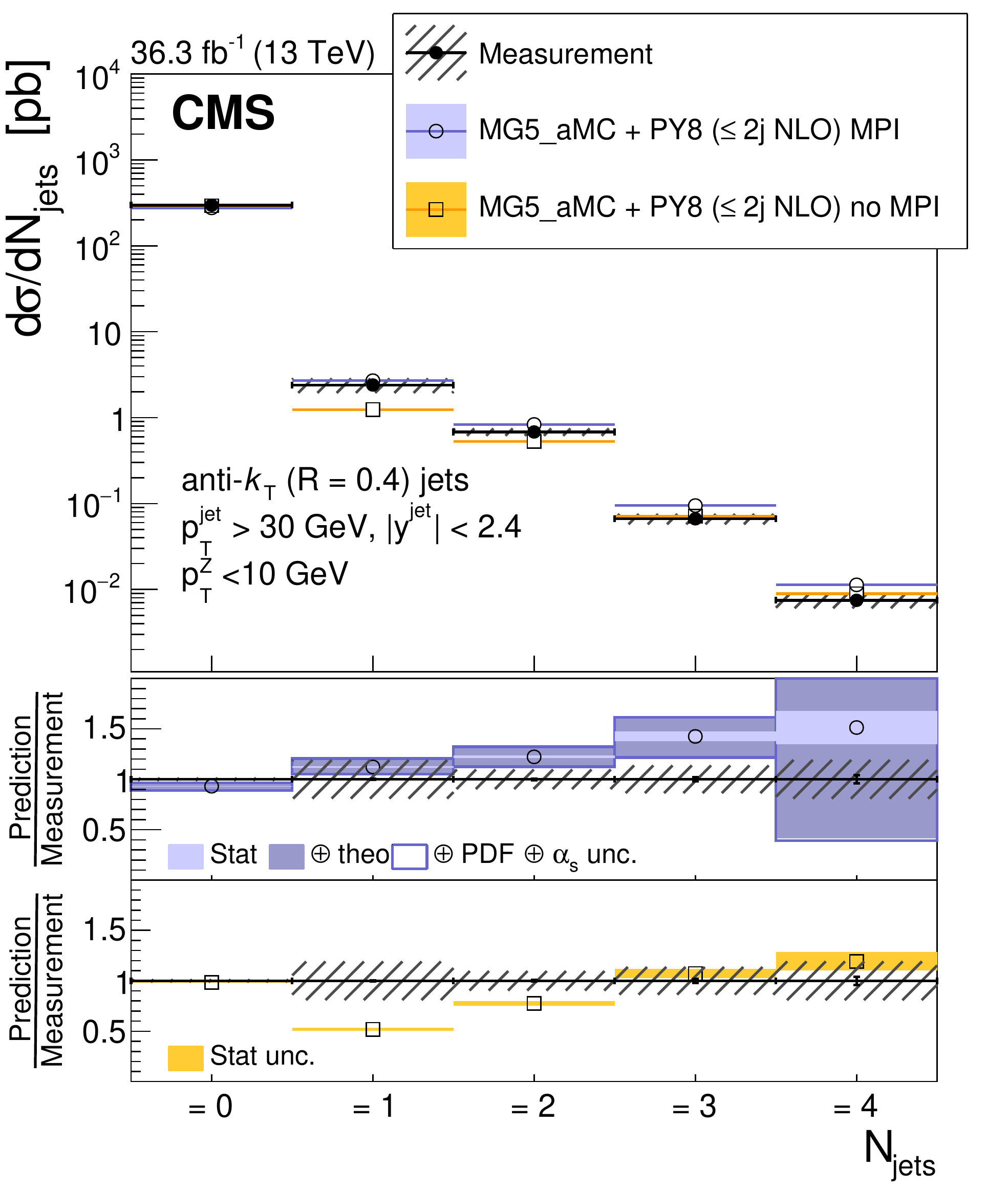}
  \includegraphics[width=0.488\textwidth]{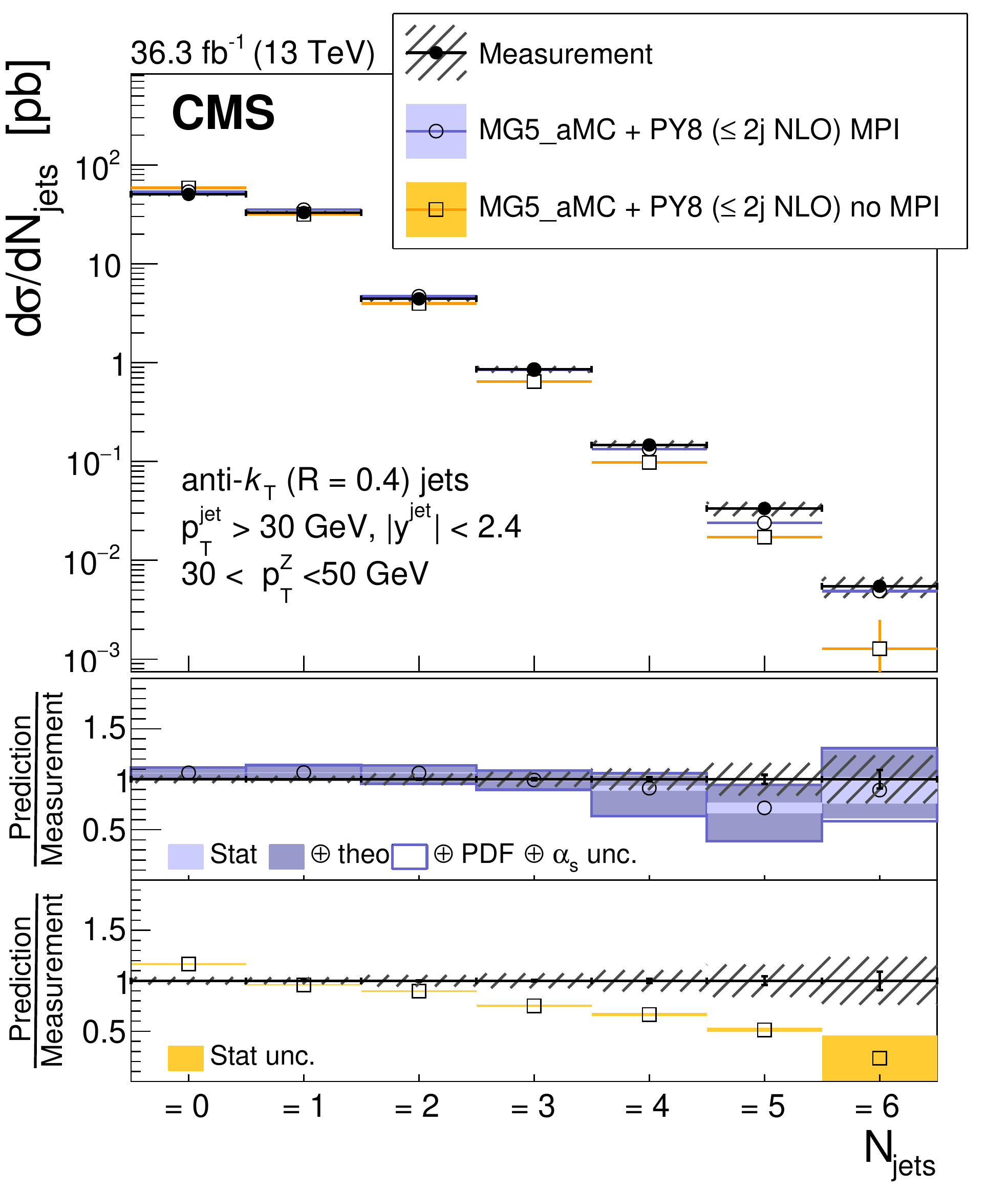}
  \includegraphics[width=0.488\textwidth]{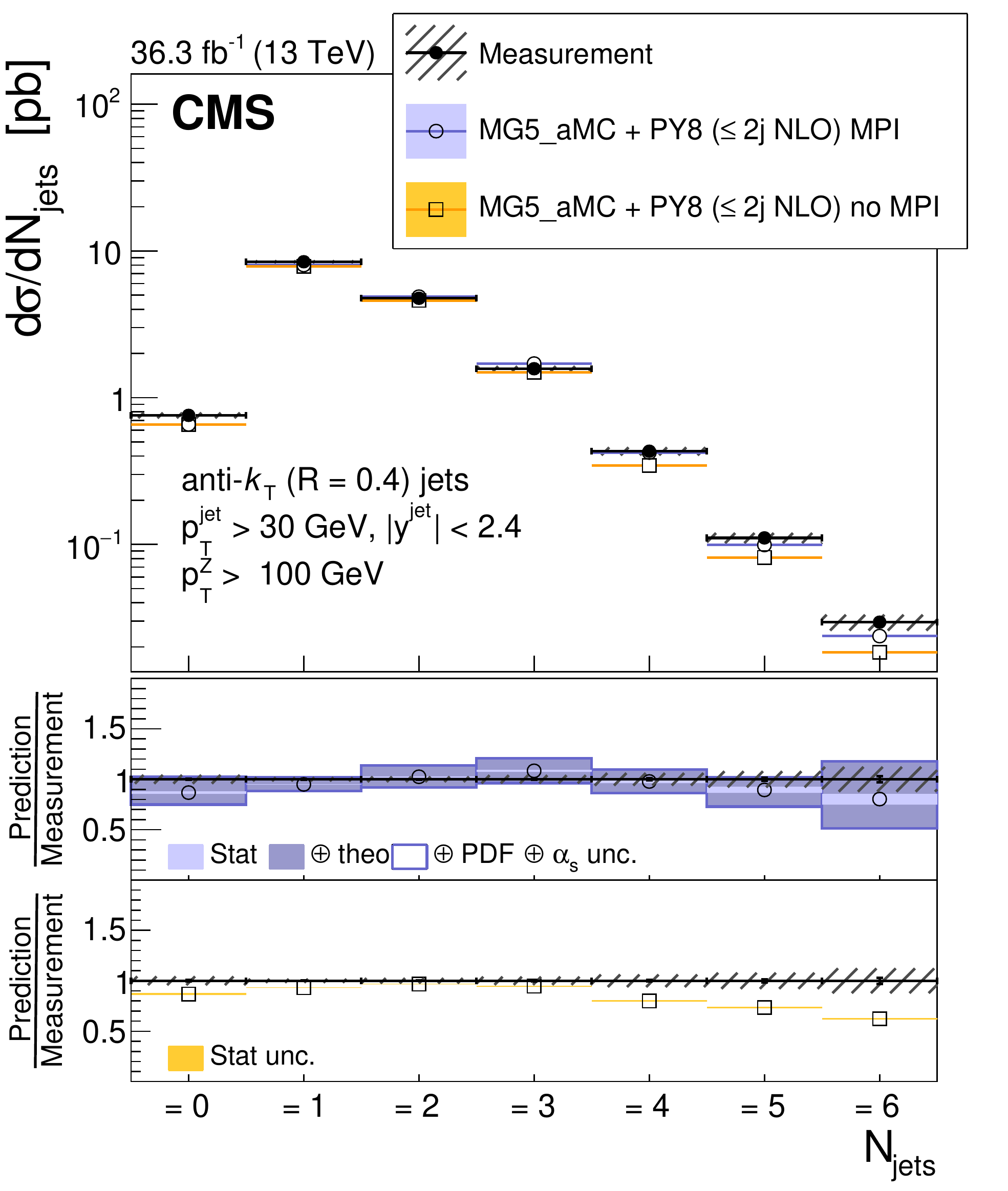}
    \topcaption{Jet multiplicity  in  three different regions of \protect\ptZ : 
 $\ptZ <10\GeV$ (upper left), $30< \ptZ < 50\GeV$ (upper right),   $\ptZ > 100\GeV $ (lower).
  The error bars on the data points represent  the statistical uncertainty of the measurement, and the hatched band shows the total statistical and systematic uncertainties added in quadrature. Predictions using \MGPyNLO with and without MPI are shown.
 }
  \label{fig:muNJ}
\end{figure*}

In Fig.~\ref{fig:muNJ}, the exclusive jet multiplicity is shown for three different ranges of \ptZ. At low \ptZ, the majority of events have no jet with $\pt > 30\GeV$ and only about 1\% of the events have one or more jets.
 This suggests that the \ptZ is mainly compensated by softer ($\pt < 30\GeV$) radiation at low \ptZ. Events with higher jet multiplicity indicate that the dominant hard process is essentially a jet production process, and the \PZ boson is radiated from a quark, as an electroweak correction to a pure QCD process.
At high \ptZ, the majority of events have at least one jet with a tail towards higher jet multiplicities, which indicates that the hardest process is indeed \PZ{}+jet, and additional jets originate from higher-order QCD corrections.

{\tolerance=1000
In Fig.~\ref{fig:muNJ}, the measurement is compared with the generator \MGPyNLO with and without multiparton interactions. 
The MPI contribution is important in the low \ptZ region, but also at higher \ptZ and higher jet multiplicities MPI plays a role.
 The prediction of \MGPyNLO including MPI agrees with the measurement, even for high jet multiplicities.
This behaviour is consistent with the prediction of the dependence of MPI effects on event kinematics from the \ptZ reported in~\cite{Bansal:2016iri}.
\par}

\begin{figure*}[htbp]
  \centering
    \includegraphics[width=0.488\textwidth]{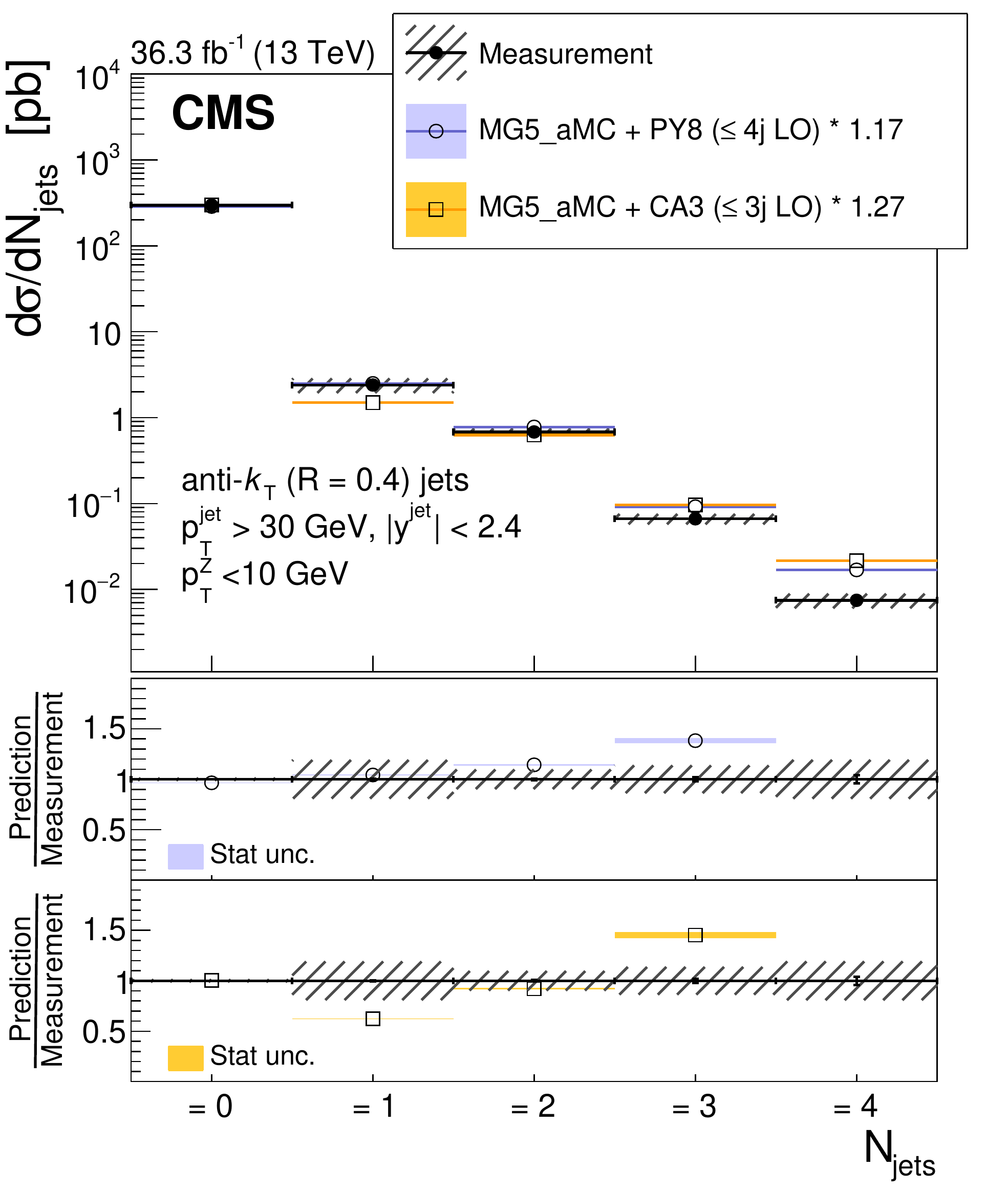}
    \includegraphics[width=0.488\textwidth]{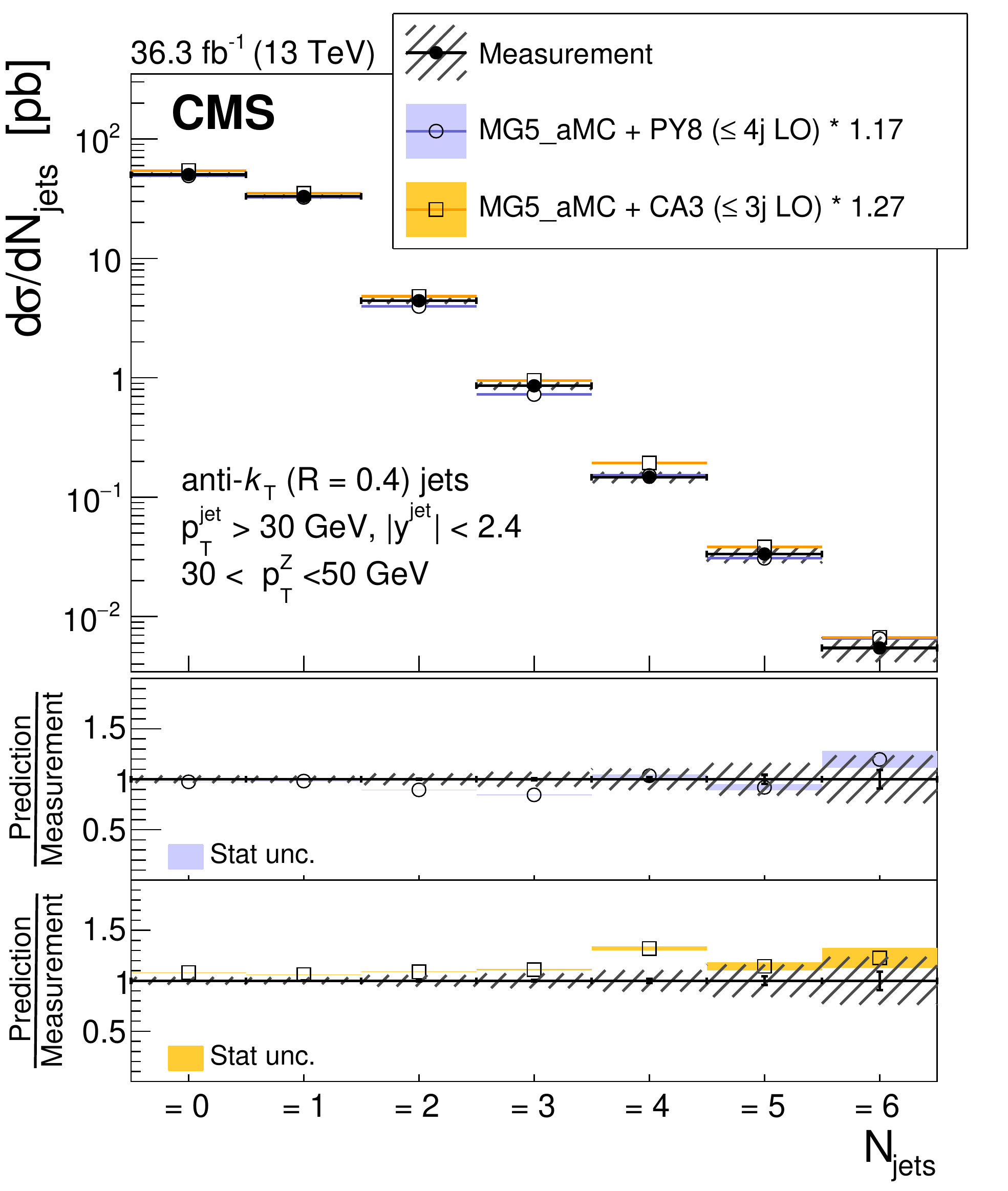}
    \includegraphics[width=0.488\textwidth]{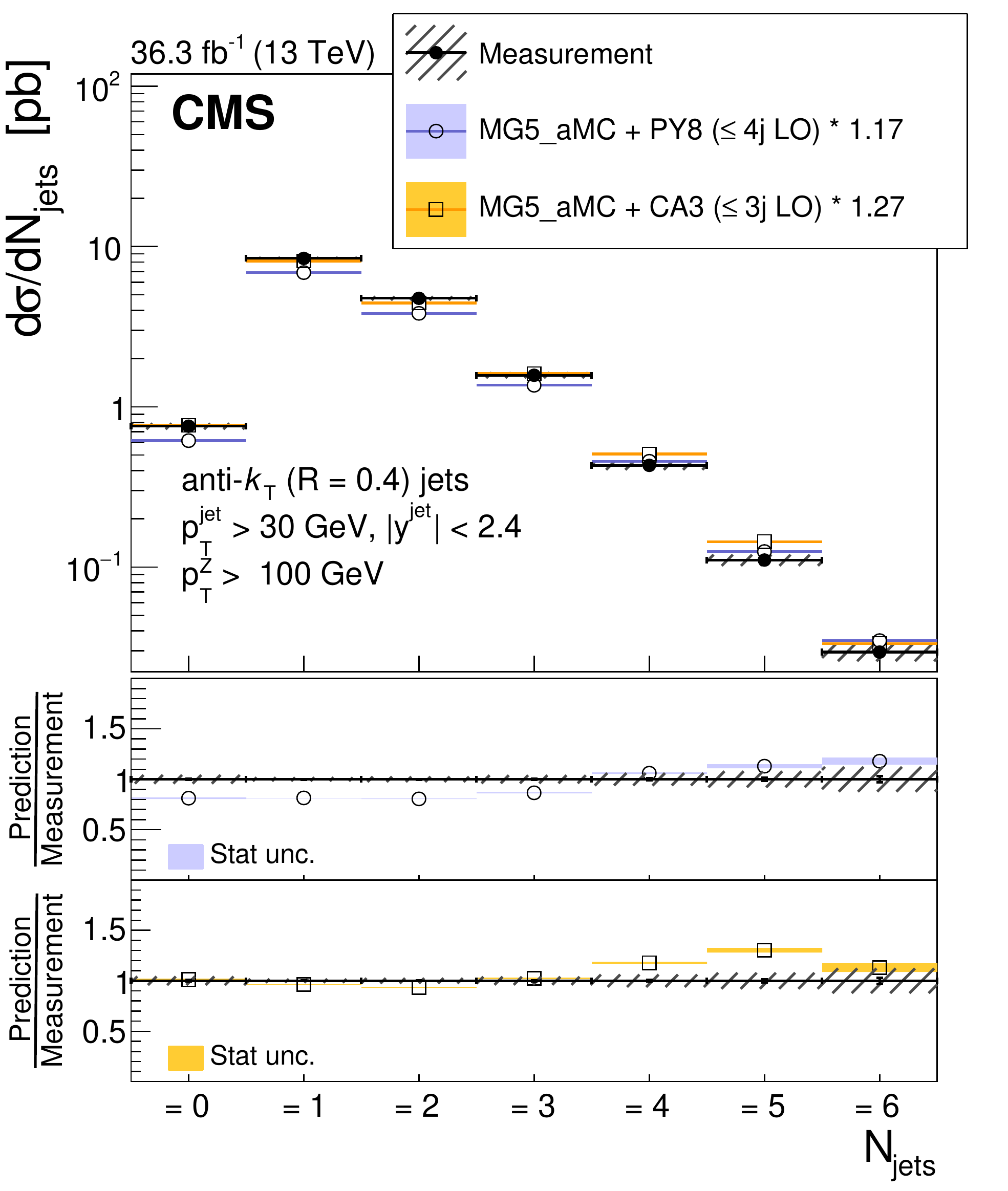}
    \topcaption{Jet multiplicity  in three different regions of \protect\ptZ : 
     $\ptZ <10\GeV$ (upper left), $30< \ptZ < 50\GeV$ (upper right),   $\ptZ > 100\GeV $ (lower).
  The error bars on the data points represent  the statistical uncertainty of the measurement, and the hatched band shows  the total statistical and systematic uncertainties added in quadrature.  
  Predictions from \MGPyLO  and \MGTMDMLM  are shown. Different normalization factors are applied, as described in the text.}
    \label{fig:muNJ-MLM}
  \end{figure*}

{\tolerance=3000
In Fig.~\ref{fig:muNJ-geneva-PB}, a comparison of the measurement with predictions from \MGTMDZj, \MGTMDZjj  and \geneva is shown. 
Both \MGTMDZj and \MGTMDZjj predictions are multiplied by a factor 1.2 to account for the normalization of  \PBM TMD~set~2 (as discussed in Section~\ref{sec:theory}).
For $\ptZ > 30\GeV$ the \PZ{}+1 (\PZ{}+2) predictions describe well the one (two) jet multiplicities, whereas at higher multiplicities a deviation from these measurement is observed, which can be attributed to the missing MPI contributions (as shown in Fig.~\ref{fig:muNJ}). The \geneva predictions, which include MPI, are in agreement for low jet multiplicities for low \ptZ, whereas higher jet multiplicities are not well described because of missing higher order contributions in the ME calculations.

\begin{figure*}[htbp]
  \centering
    \includegraphics[width=0.488\textwidth]{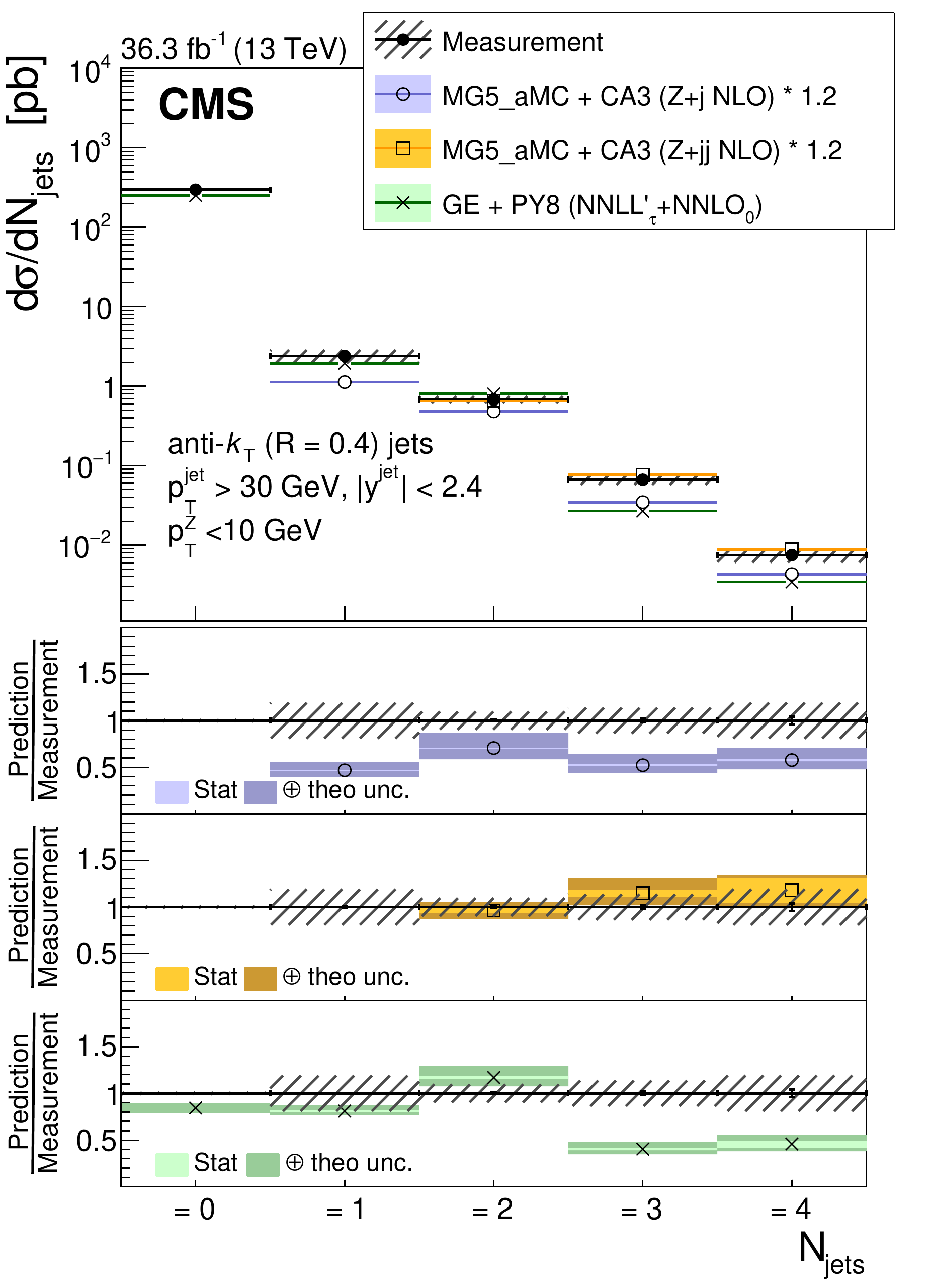}
    \includegraphics[width=0.488\textwidth]{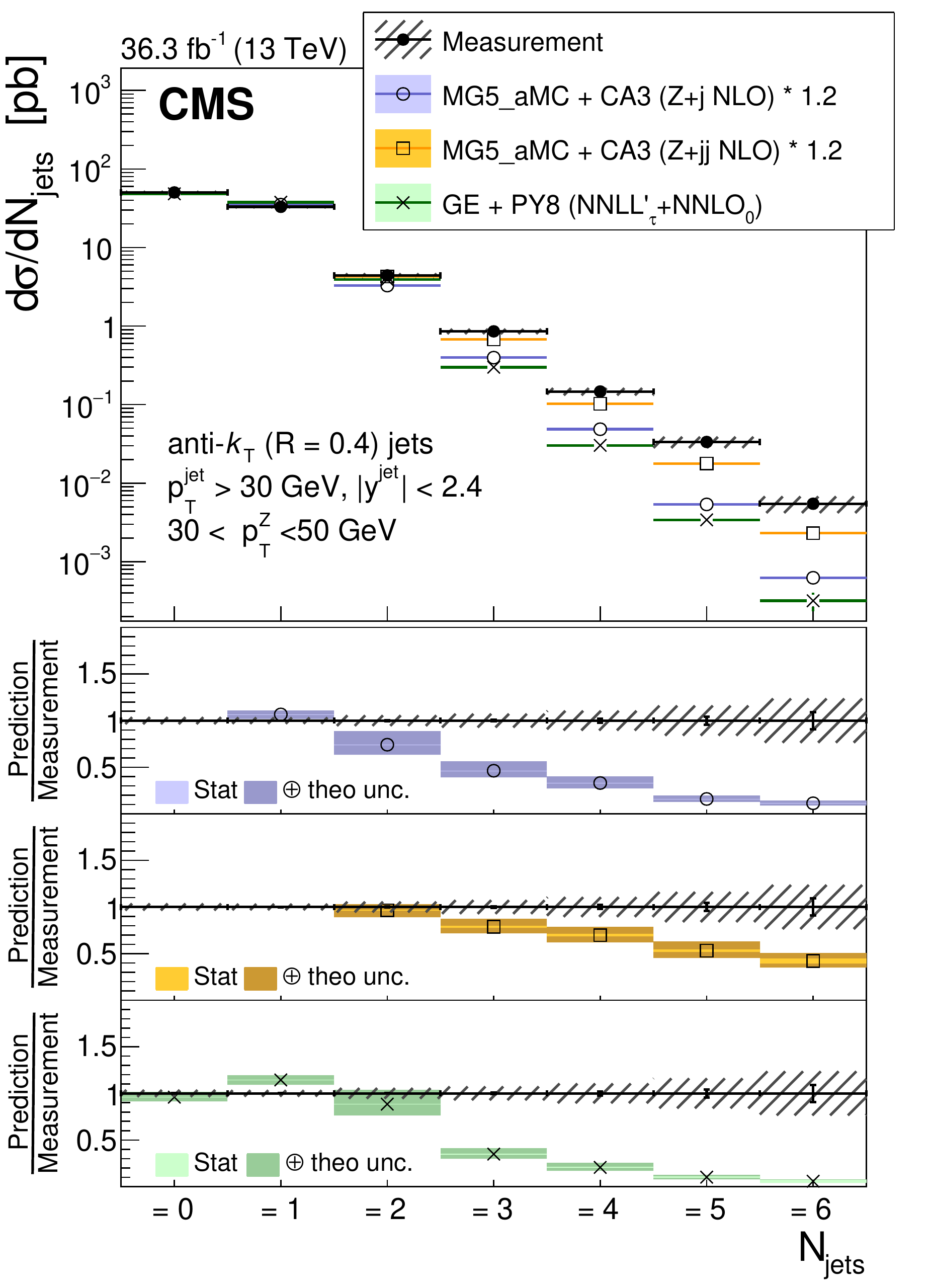}
    \includegraphics[width=0.488\textwidth]{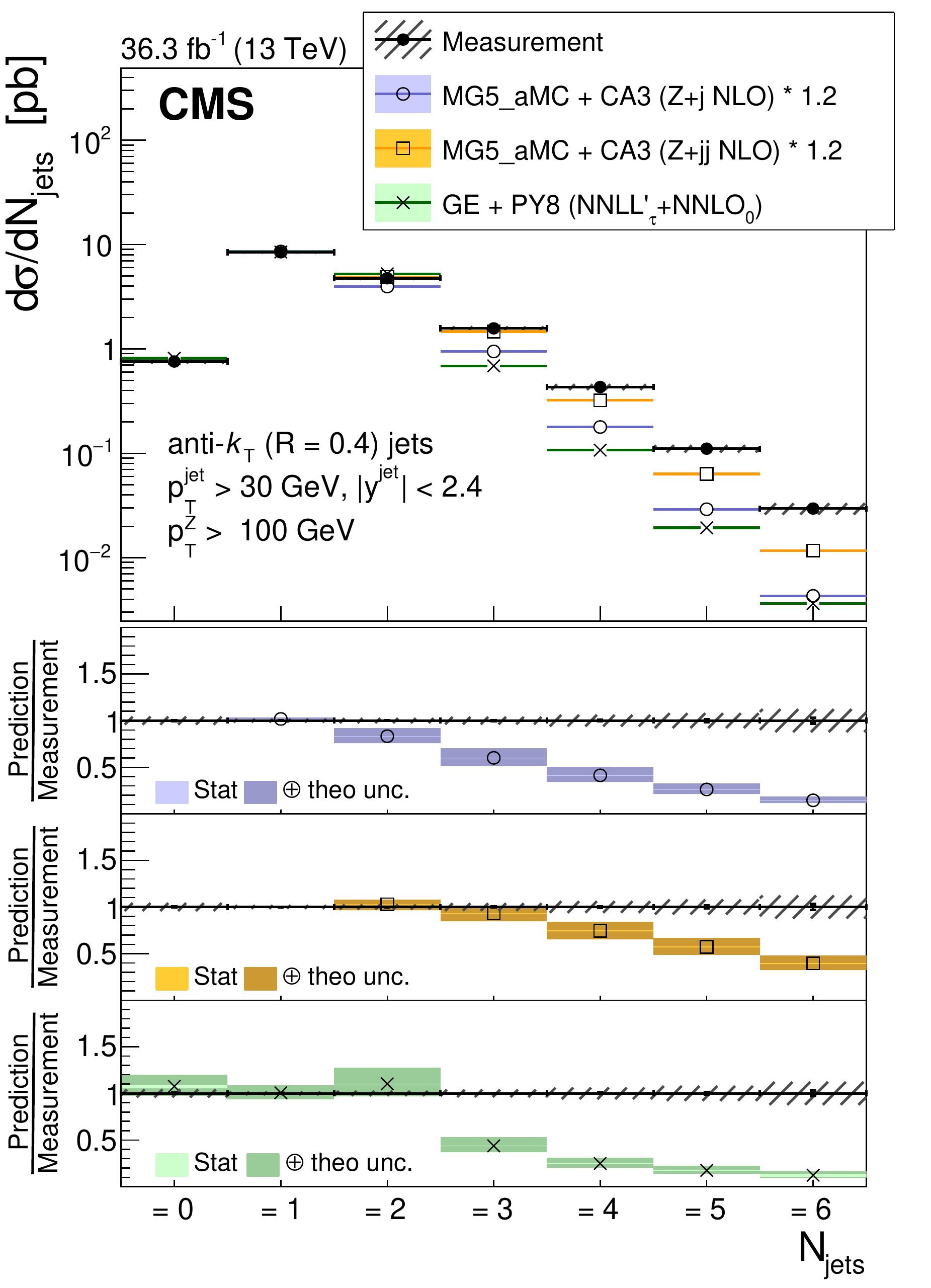}
    \topcaption{Jet multiplicity  in three different regions of \protect\ptZ : 
    $\ptZ <10\GeV$ (upper left), $30< \ptZ < 50\GeV$ (upper right),   $\ptZ > 100\GeV $ (lower).
  The error bars on the data points represent  the statistical uncertainty of the measurement, and the hatched band shows  the total statistical and systematic uncertainties added in quadrature.  
  Predictions from \MGTMDZj, \MGTMDZjj and  \geneva are shown. An overall normalization factor of 1.2 is applied to \MGTMDZj  and \MGTMDZjj.}
    \label{fig:muNJ-geneva-PB}
  \end{figure*}

 In Fig.~\ref{fig:muNJ-MLM}, the measurement is compared with predictions from \MGPyLO  and \MGTMDMLM.  
The prediction from \MGPyLO describes the measurements in all \ptZ regions.
  The \MGTMDMLM prediction agrees with the measurements in all \ptZ ranges, except in the second bin at low \ptZ values where MPI plays a significant role.
\par}

\begin{figure*}[htbp]
  \centering
    \includegraphics[width=0.488\textwidth]{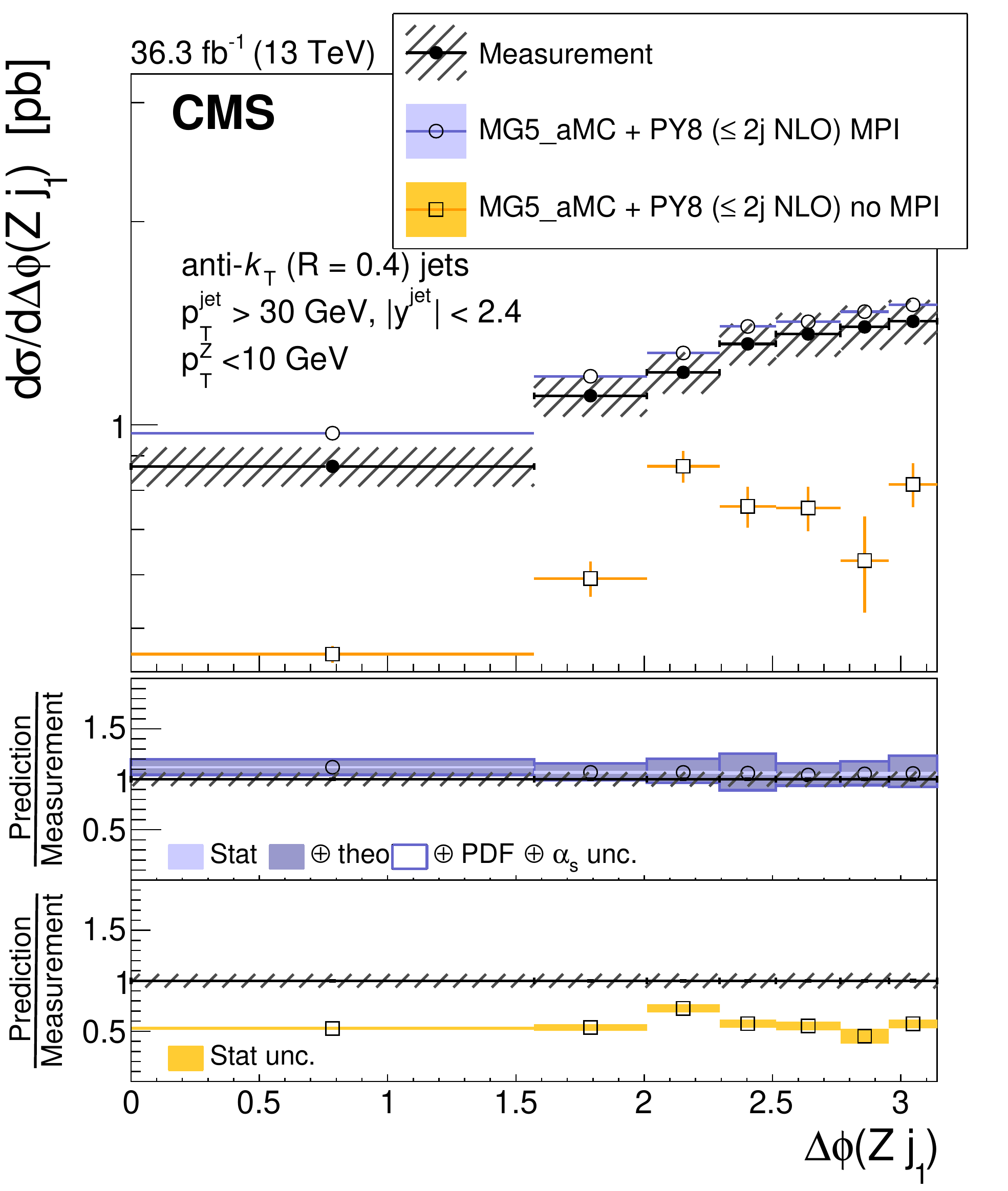}
    \includegraphics[width=0.488\textwidth]{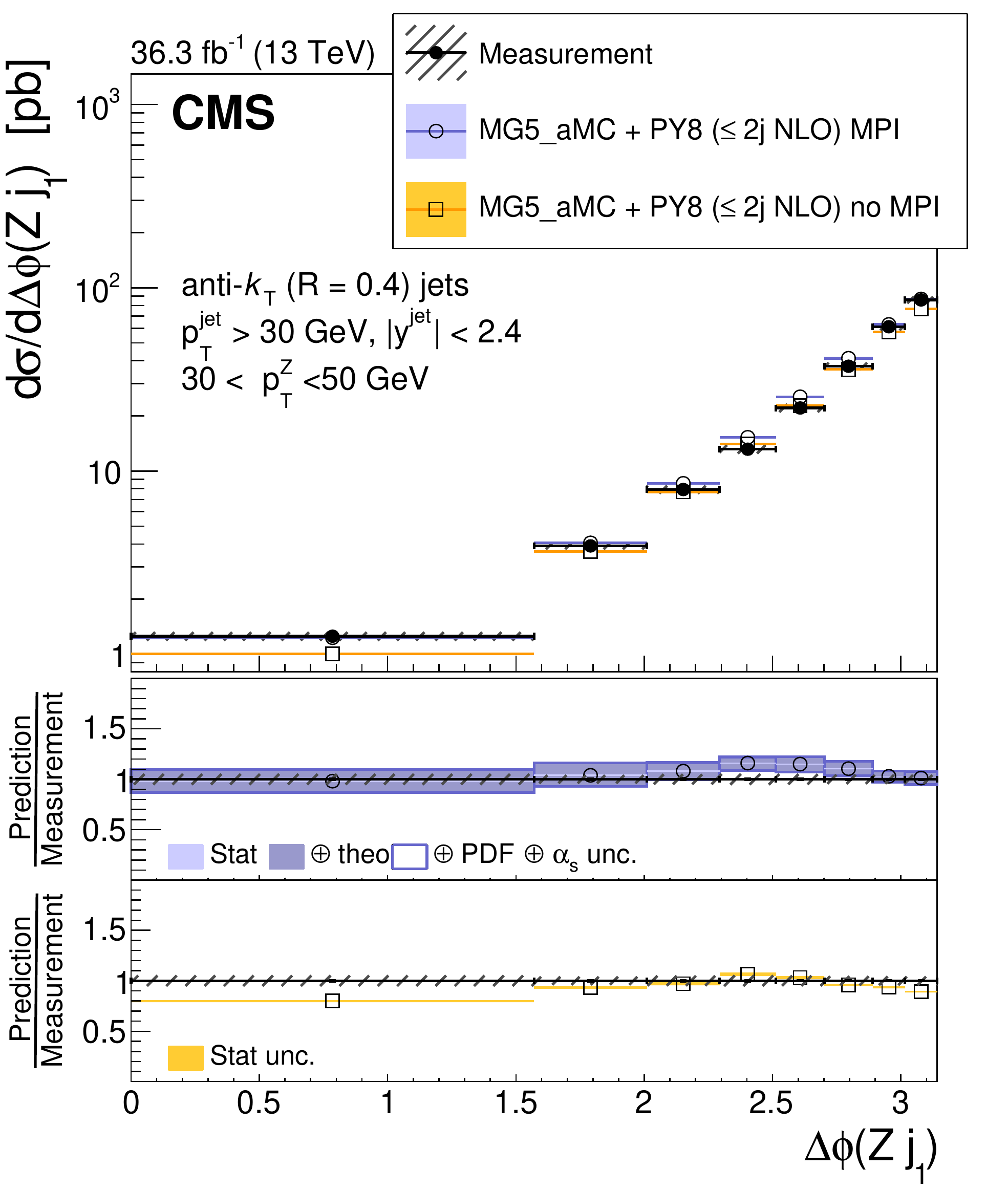}
    \includegraphics[width=0.488\textwidth]{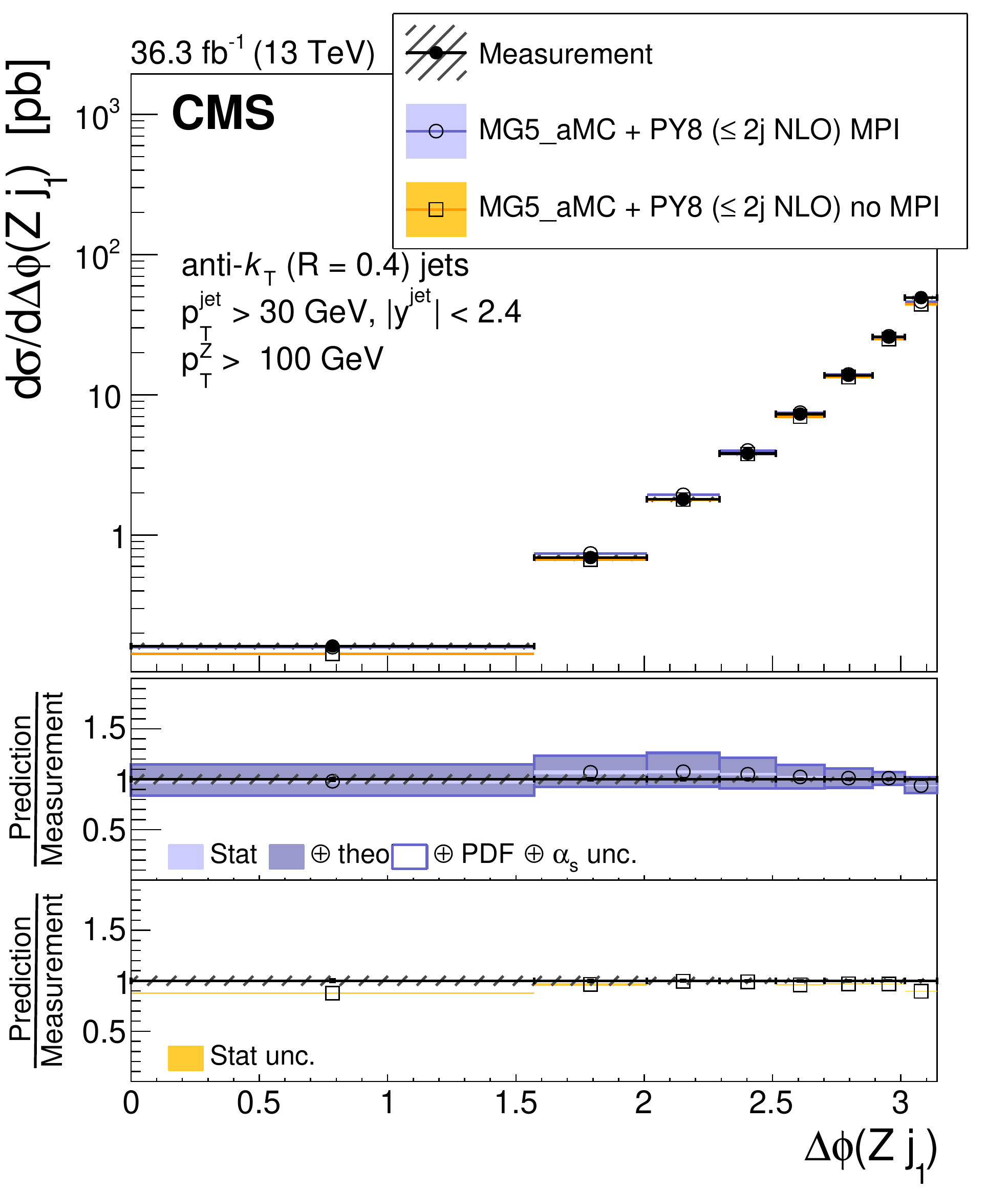}
      \topcaption{Cross section as a function of \dphiZj between the \PZ\ boson and the leading jet in the three \ptZ\ bins:
     $\ptZ <10\GeV$ (upper left), $30< \ptZ < 50\GeV$ (upper right),   $\ptZ > 100\GeV $ (lower).
  The error bars on the data points represent  the statistical uncertainty of the measurement, and the hatched band shows  the total statistical and systematic uncertainties added in quadrature.  Predictions using \MGPyNLO with and without multi-parton interactions are shown.
   }
    \label{fig:muDPhiZJ}
  \end{figure*}

In Fig.~\ref{fig:muDPhiZJ}, the azimuthal correlation, \dphiZj, between the \PZ boson and the leading jet is shown for three different ranges of \ptZ. In the range $\ptZ < 10\GeV$, the \PZ boson is only very weakly correlated with the leading jet, thus the distribution is almost uniform. In the region $ \ptZ > 100\GeV $, the \PZ boson is highly correlated with the leading jet and the cross section 
falls more than two orders of magnitude from the back-to-back region to the small \dphiZj region. The systematic uncertainty in the low \ptZ range is $\mathcal{O}(10 \%)$. 
 In Fig.~\ref{fig:muDPhiZJ}, the predictions from \MGPyNLO with and without MPI are compared with the measurement. In the low \ptZ range,  MPI contributes about 40\%, and
 even in the region of $ 30 < \ptZ < 50\GeV$, the contribution from MPI could be about 20\%  in the small-\dphiZj region. The prediction of \MGPyNLO including MPI describes the measurements.
 
\begin{figure*}[htbp]
 \centering
  \includegraphics[width=0.488\textwidth]{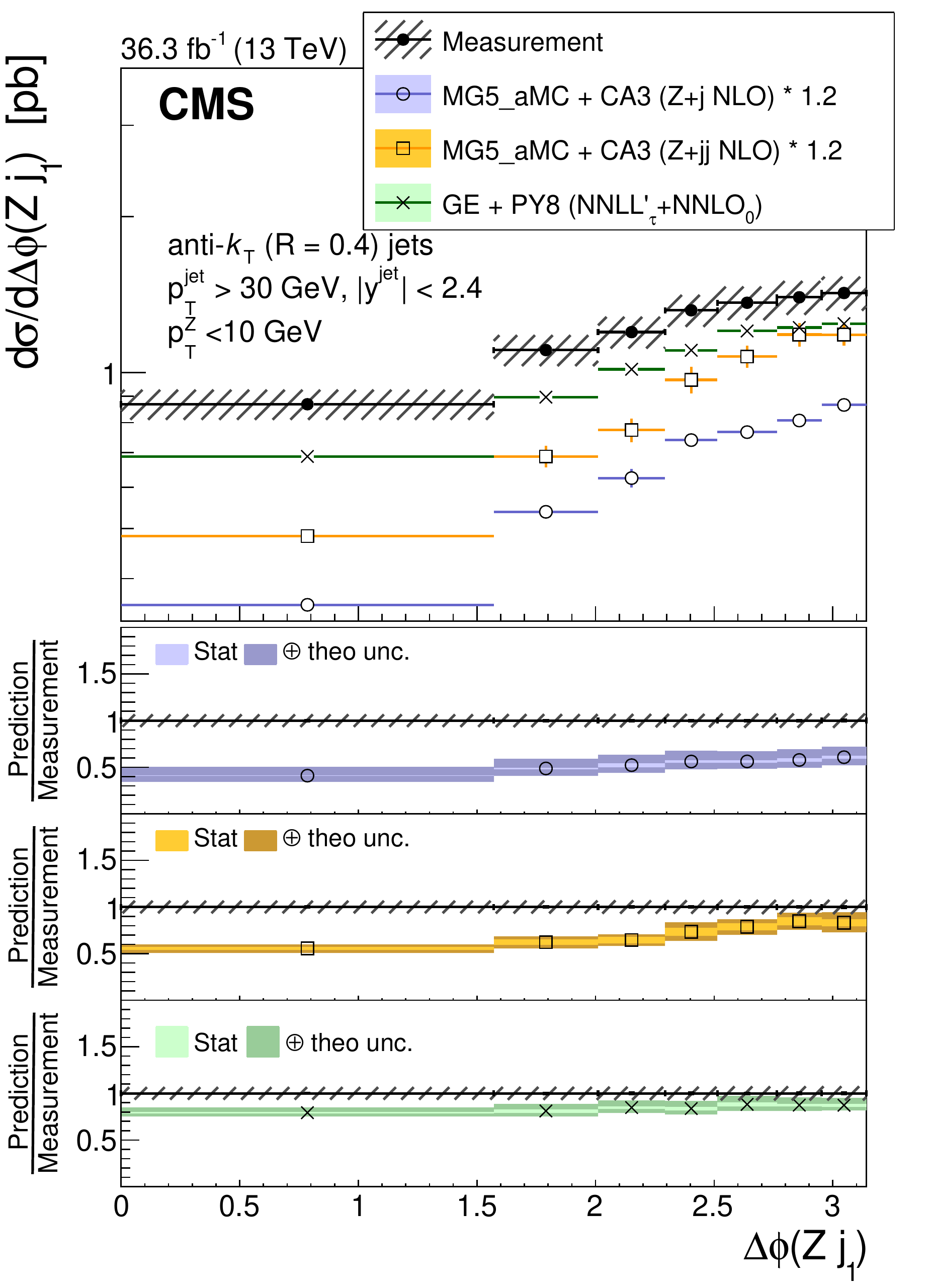}
  \includegraphics[width=0.488\textwidth]{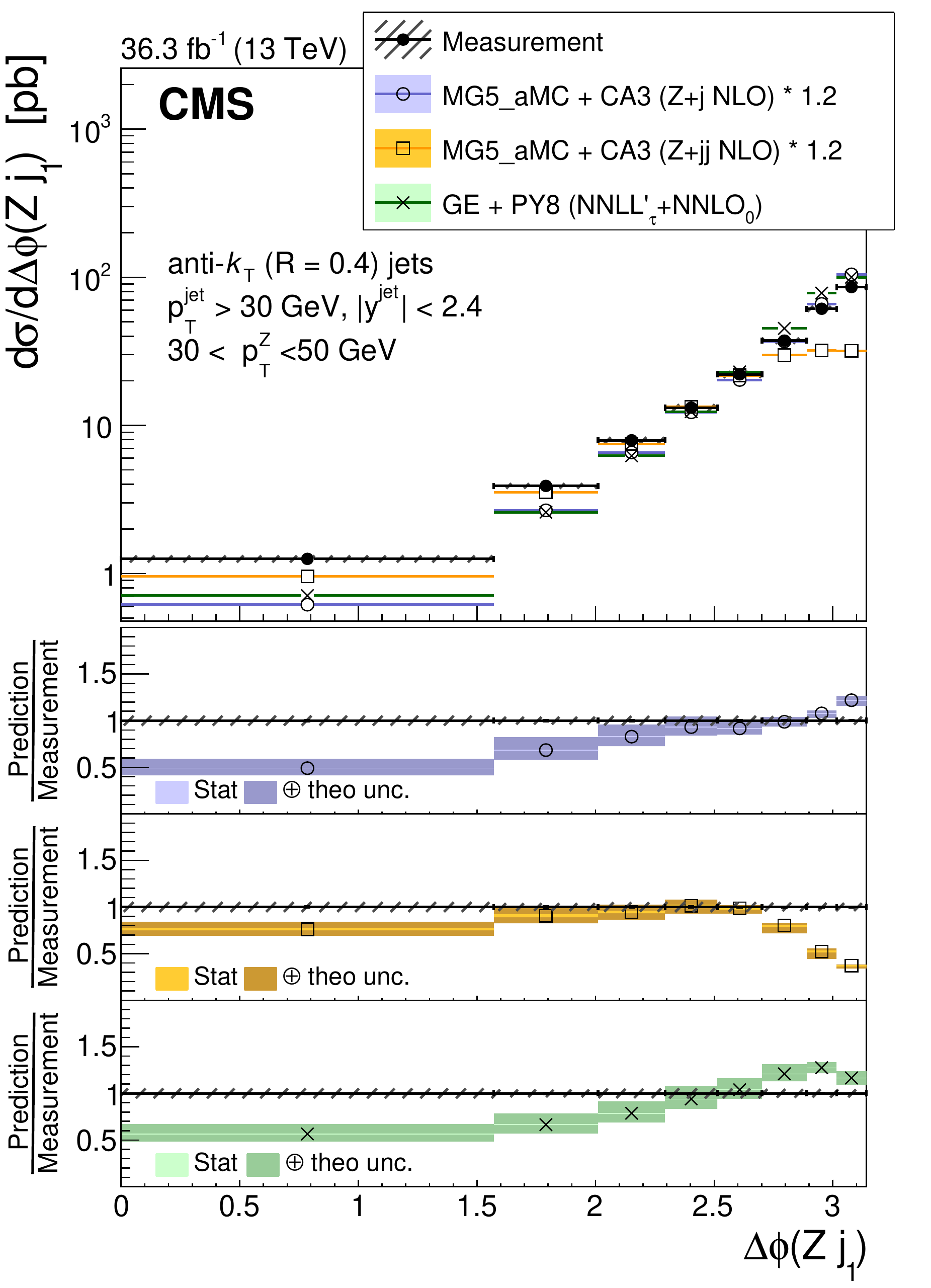}
  \includegraphics[width=0.488\textwidth]{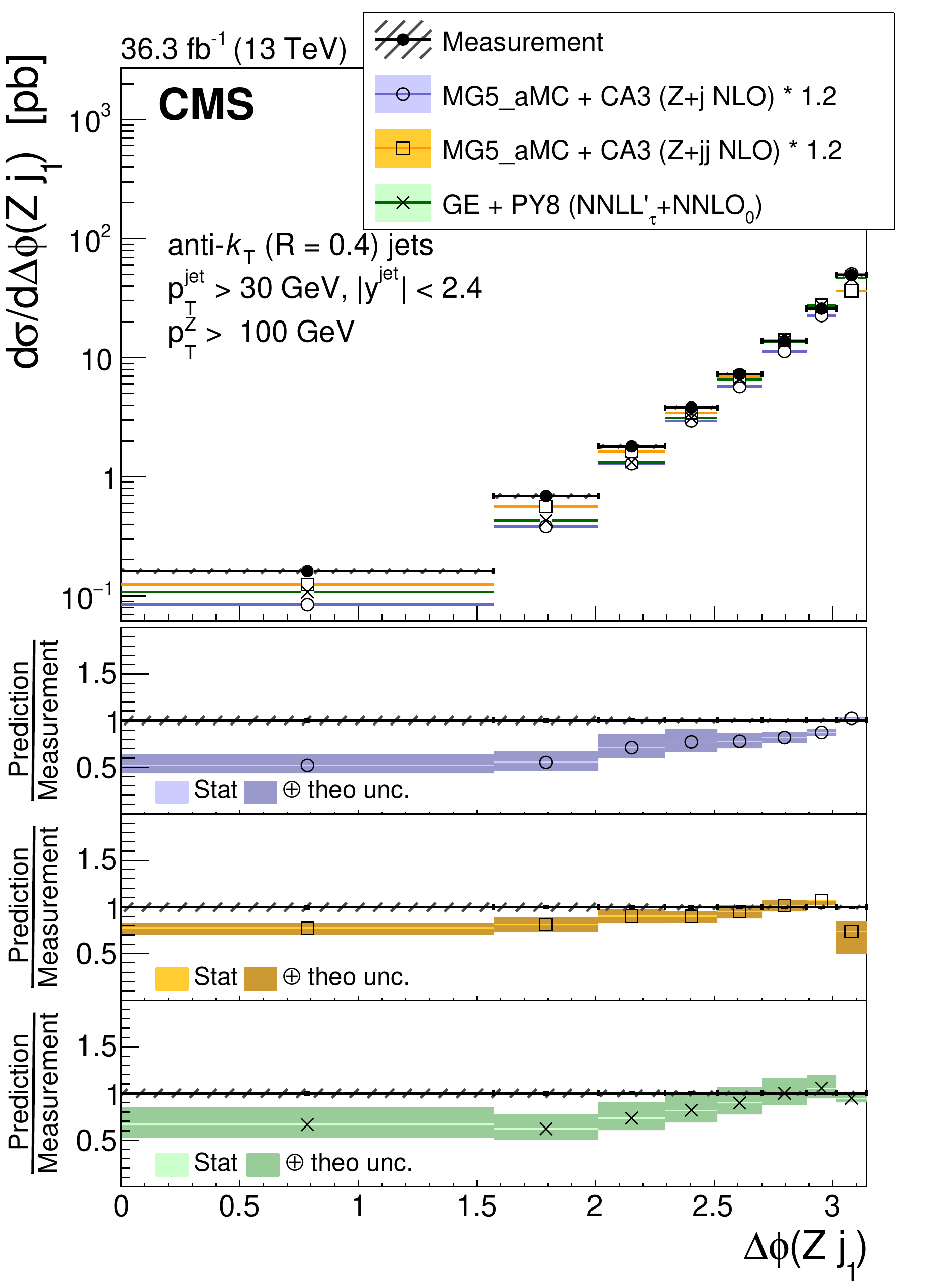}
  \topcaption{Cross section as a function of \dphiZj between the \PZ\ boson and the leading jet in the three \ptZ\ bins:
   $\ptZ <10\GeV$ (upper left), $30< \ptZ < 50\GeV$ (upper right),   $\ptZ > 100\GeV $ (lower).
The error bars on the data points represent  the statistical uncertainty of the measurement, and the hatched band shows  the total statistical and systematic uncertainties added in quadrature.  
Predictions from \geneva, \MGTMDZj  and \MGTMDZjj  are shown. An overall normalization factor of 1.2 is applied to \MGTMDZj  and \MGTMDZjj.}
  \label{fig:muDPhiZJ-geneva-PB}
\end{figure*}

{\tolerance=3000
 In Fig.~\ref{fig:muDPhiZJ-geneva-PB}, the measurement is compared with  \MGTMDZj, \MGTMDZjj  and \textsc{Geneva (\PZ{}+0 NNLO)}. In low \ptZ range,   the \MGTMDZj  and \MGTMDZjj predictions differ from the measurements due to the missing contribution of MPI. In the high \ptZ region the predictions agree better with the measurements (the region $\dphiZj \to \pi$ is not accessible in the \PZ{}+2 calculation). The \geneva prediction agrees with the measurement at low \ptZ, whereas at larger \ptZ the prediction differs from the measurement because of missing higher order contributions.
\par}

\begin{figure*}[htbp]
  \centering
    \includegraphics[width=0.488\textwidth]{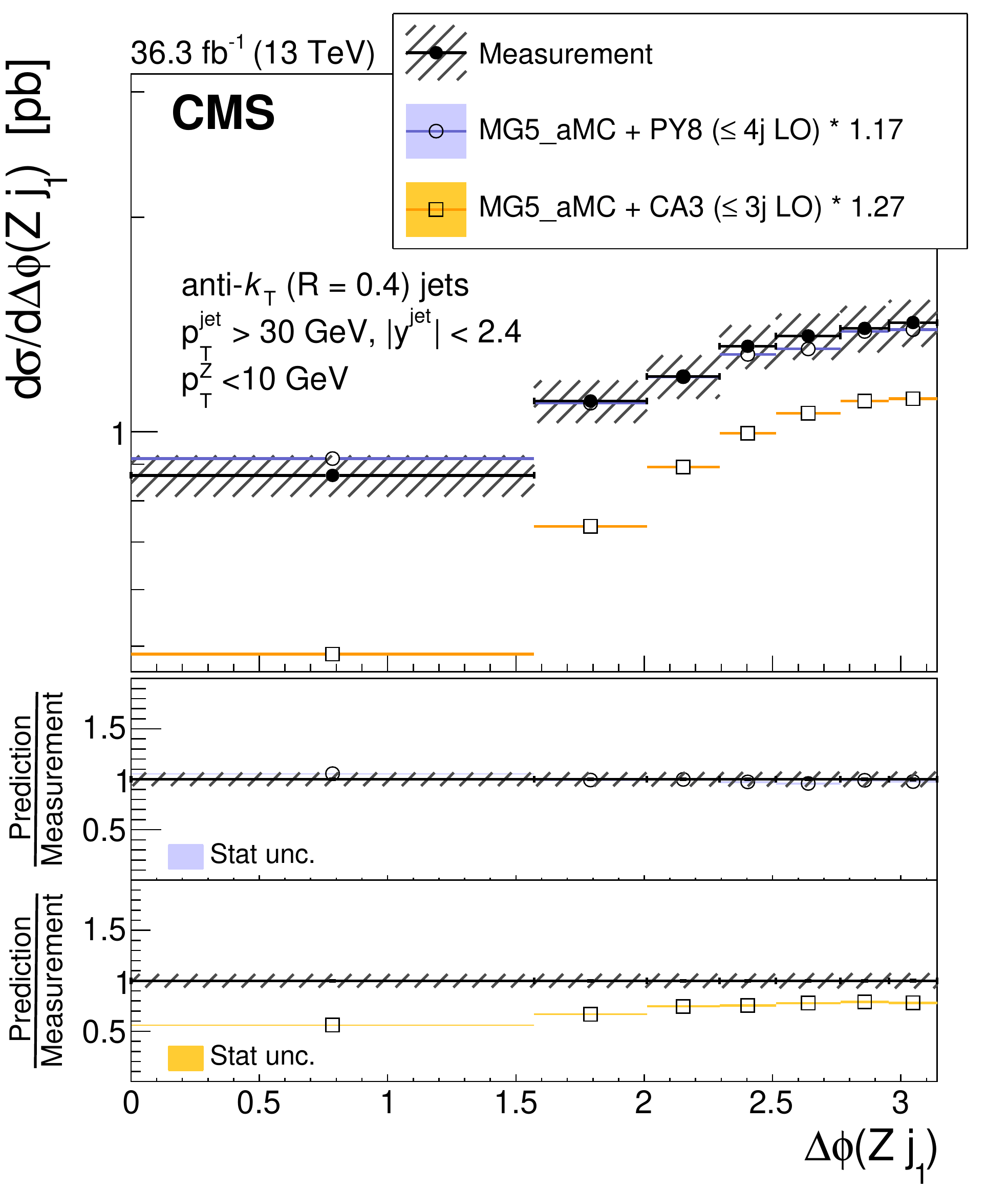}
    \includegraphics[width=0.488\textwidth]{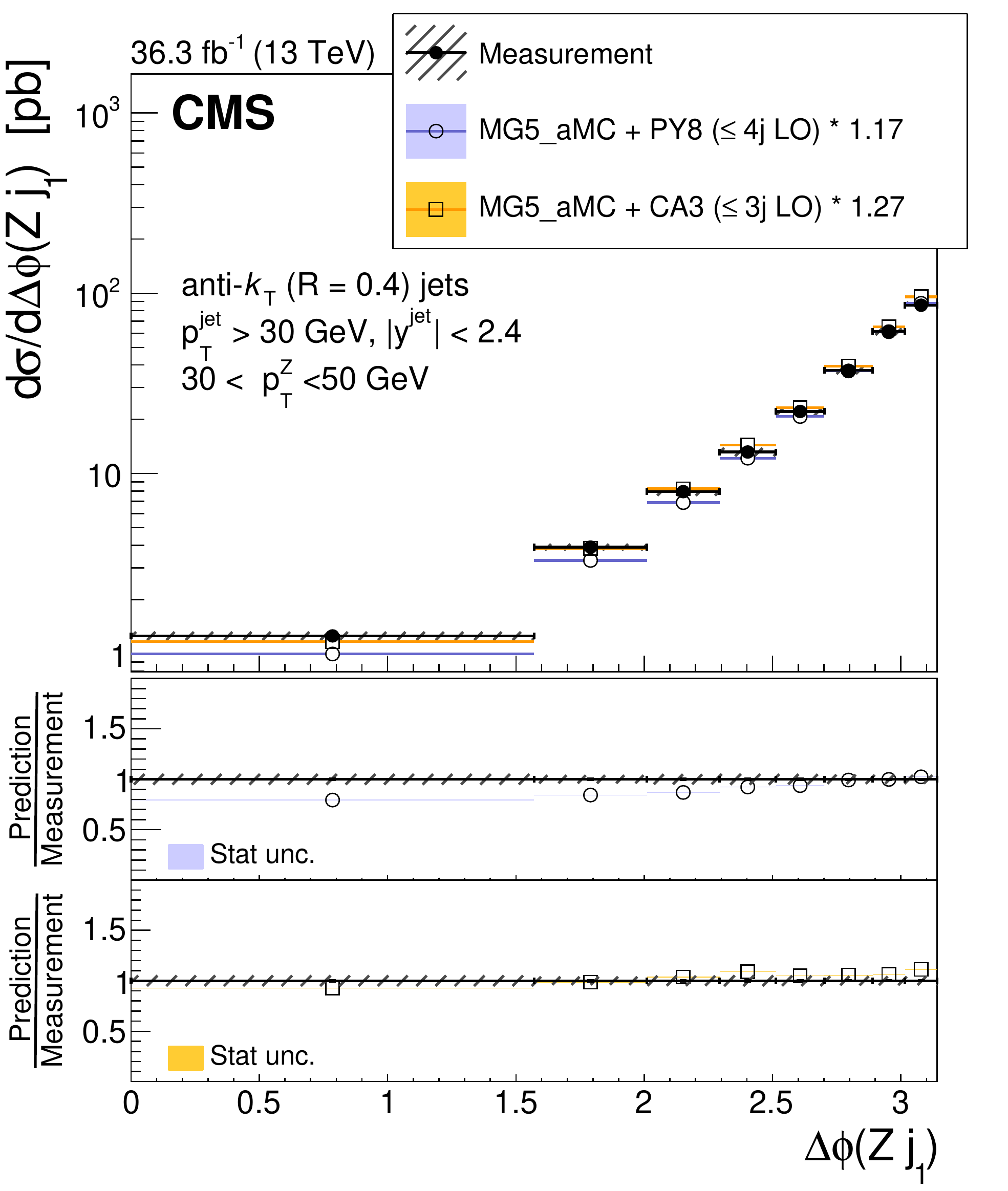}
    \includegraphics[width=0.488\textwidth]{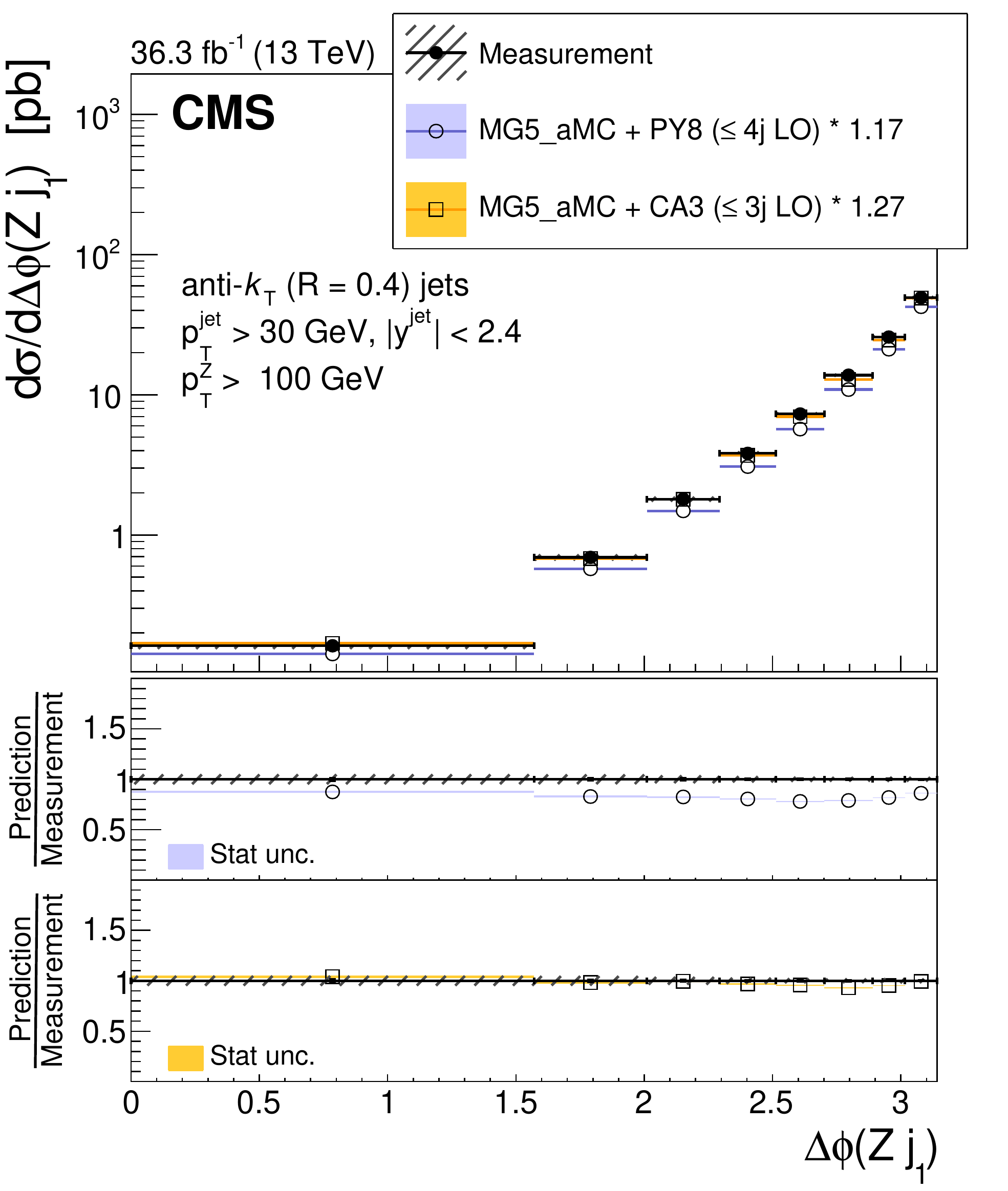}
    \topcaption{Cross section as a function of \dphiZj between the \PZ\ boson and the leading jet in three \ptZ\ bins:
    $\ptZ <10\GeV$ (upper left), $30< \ptZ < 50\GeV$ (upper right),   $\ptZ > 100\GeV $ (lower).
  The error bars on the data points represent  the statistical uncertainty of the measurement, and the hatched band shows  the total statistical and systematic uncertainties added in quadrature.  
  Predictions from \MGPyLO  and \MGTMDMLM  are shown. Different normalization factors are applied, as described in the text.}
    \label{fig:muDPhiZJ-MLM}
  \end{figure*}

In Fig.~\ref{fig:muDPhiZJ-MLM}, the predictions from \MGPyLO and \MGTMDMLM are compared with the measurement. 
 The \MGPyLO is in agreement with the measurement.
 The \MGTMDMLM prediction is too low in 
the low \ptZ region, due to the missing MPI contribution, whereas other \ptZ ranges are described.
 
\begin{figure*}[htbp]
  \centering
    \includegraphics[width=0.488\textwidth]{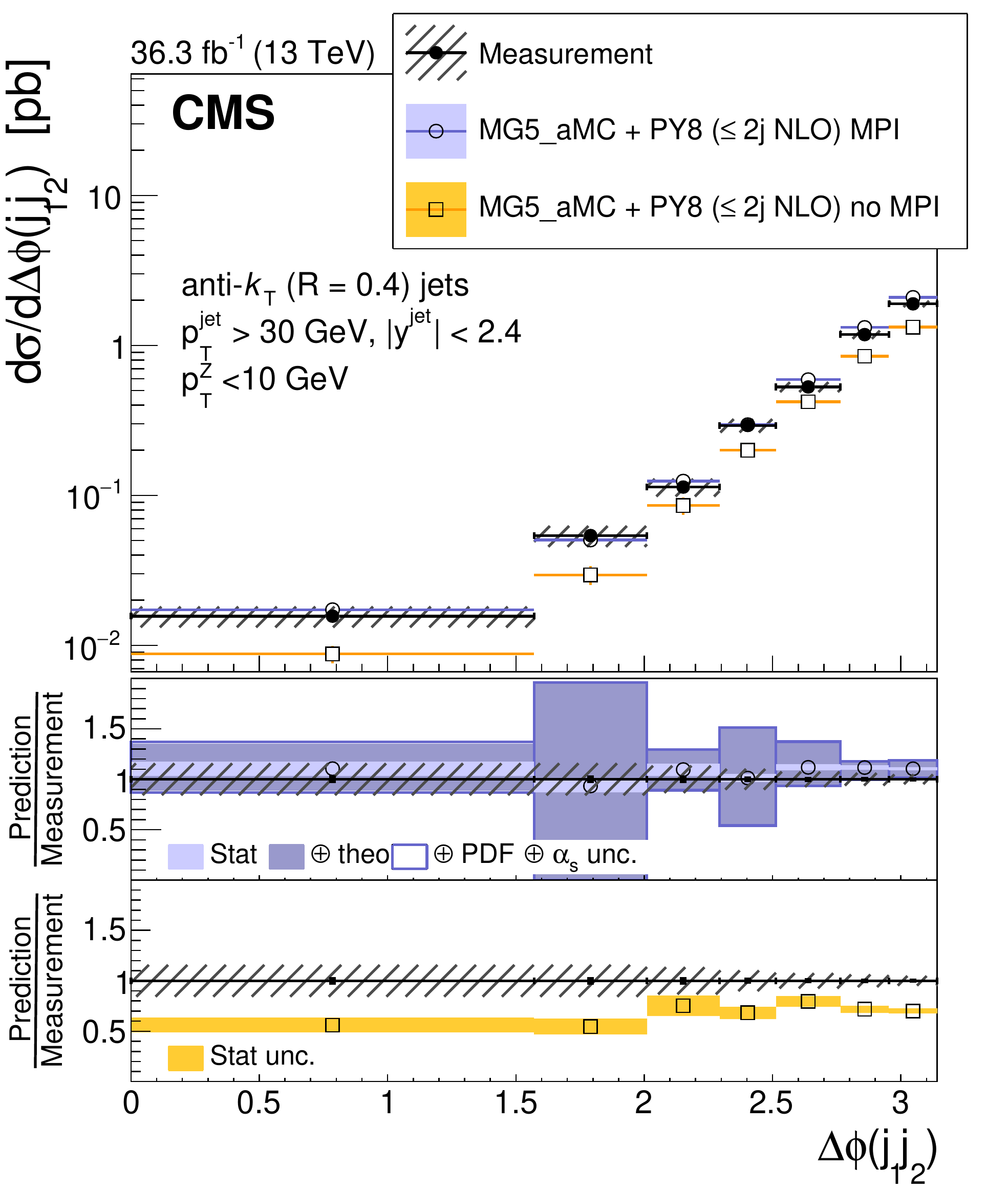}
    \includegraphics[width=0.488\textwidth]{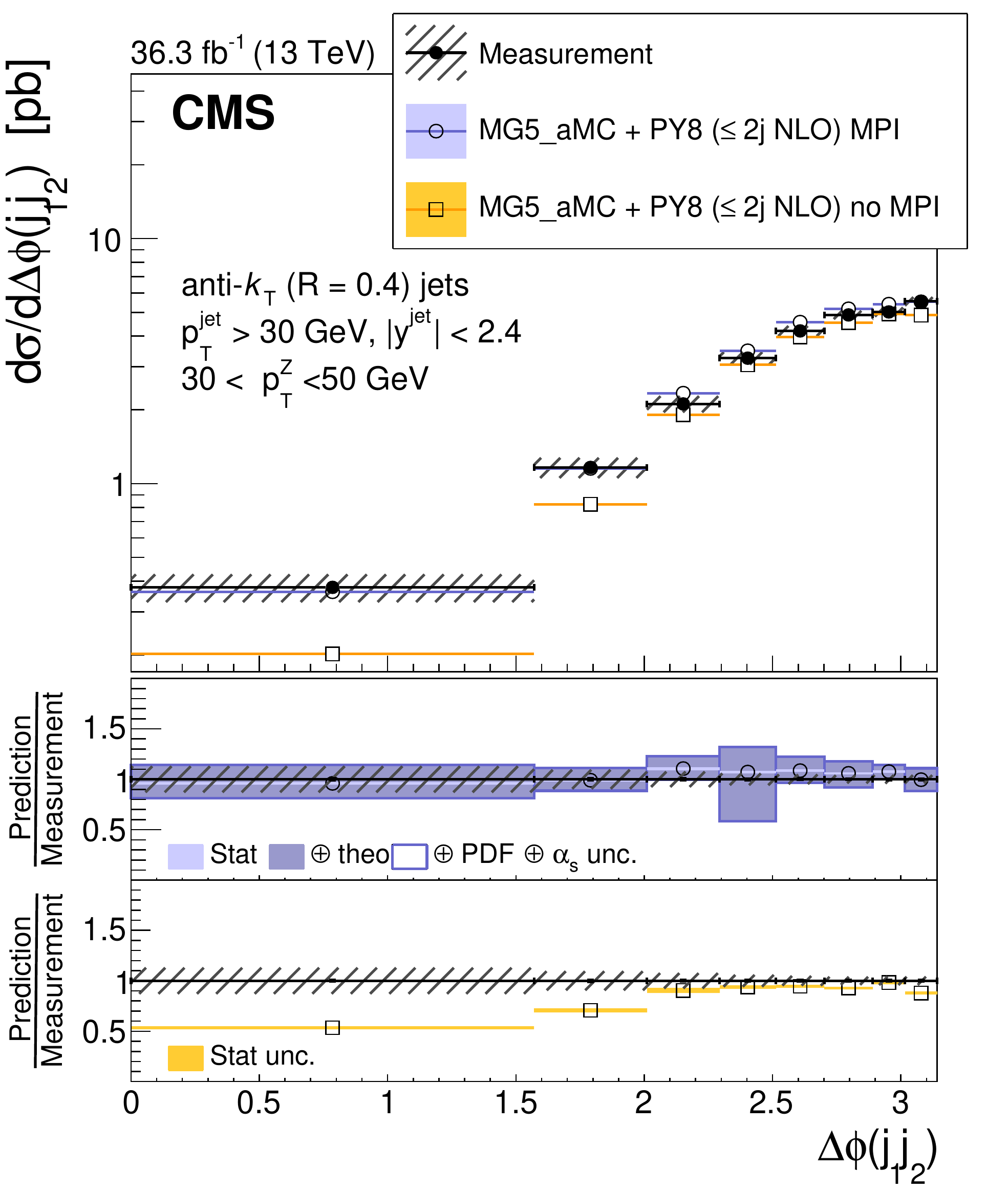}
    \includegraphics[width=0.488\textwidth]{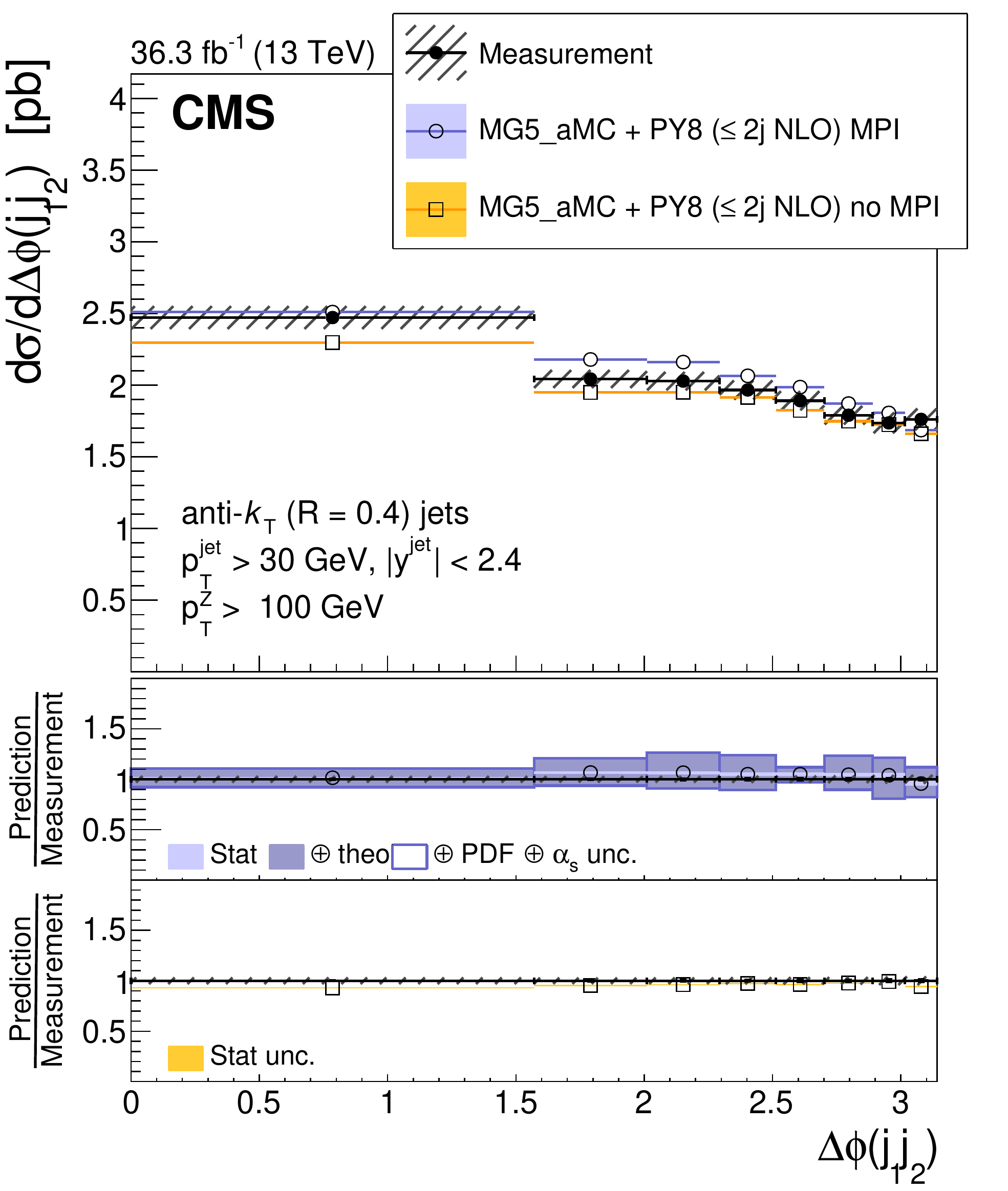}
    \topcaption{Cross section as a function of \dphiOneTwo between two leading jets in three \ptZ regions:
    $\ptZ <10\GeV$ (upper left), $30< \ptZ < 50\GeV$ (upper right),   $\ptZ > 100\GeV $ (lower).
   The error bars on the data points represent  the statistical uncertainty of the measurement, and the hatched band shows  the total statistical and systematic uncertainties added in quadrature. 
  Predictions using \MGPyNLO with and without multiparton interactions are shown.
    }
    \label{fig:muDPhiJJ}
  \end{figure*}

In Fig.~\ref{fig:muDPhiJJ}, the azimuthal correlation \dphiOneTwo between the two leading jets is shown for the three different ranges of \ptZ. A strong correlation between the two leading jets is observed at small \ptZ, whereas only a weak correlation is seen at large \ptZ. This indicates that at low \ptZ  the process is dominated by a jet production process and that 
the \PZ boson is radiated from a quark (EW correction) and therefore the jets are correlated.  On the contrary, at large \ptZ the process is dominated by \PZ{}+jet production, with higher-order QCD corrections in form of additional jets, which are only weakly correlated. 
The measurement is compared with \MGPyNLO, with and without MPI. Except for the highest \ptZ region, the contribution from MPI is significant, especially in the small \dphiOneTwo range. The prediction obtained with \MGPyNLO including MPI describes the measurement well over the whole range.

\begin{figure*}[htbp]
  \centering
    \includegraphics[width=0.488\textwidth]{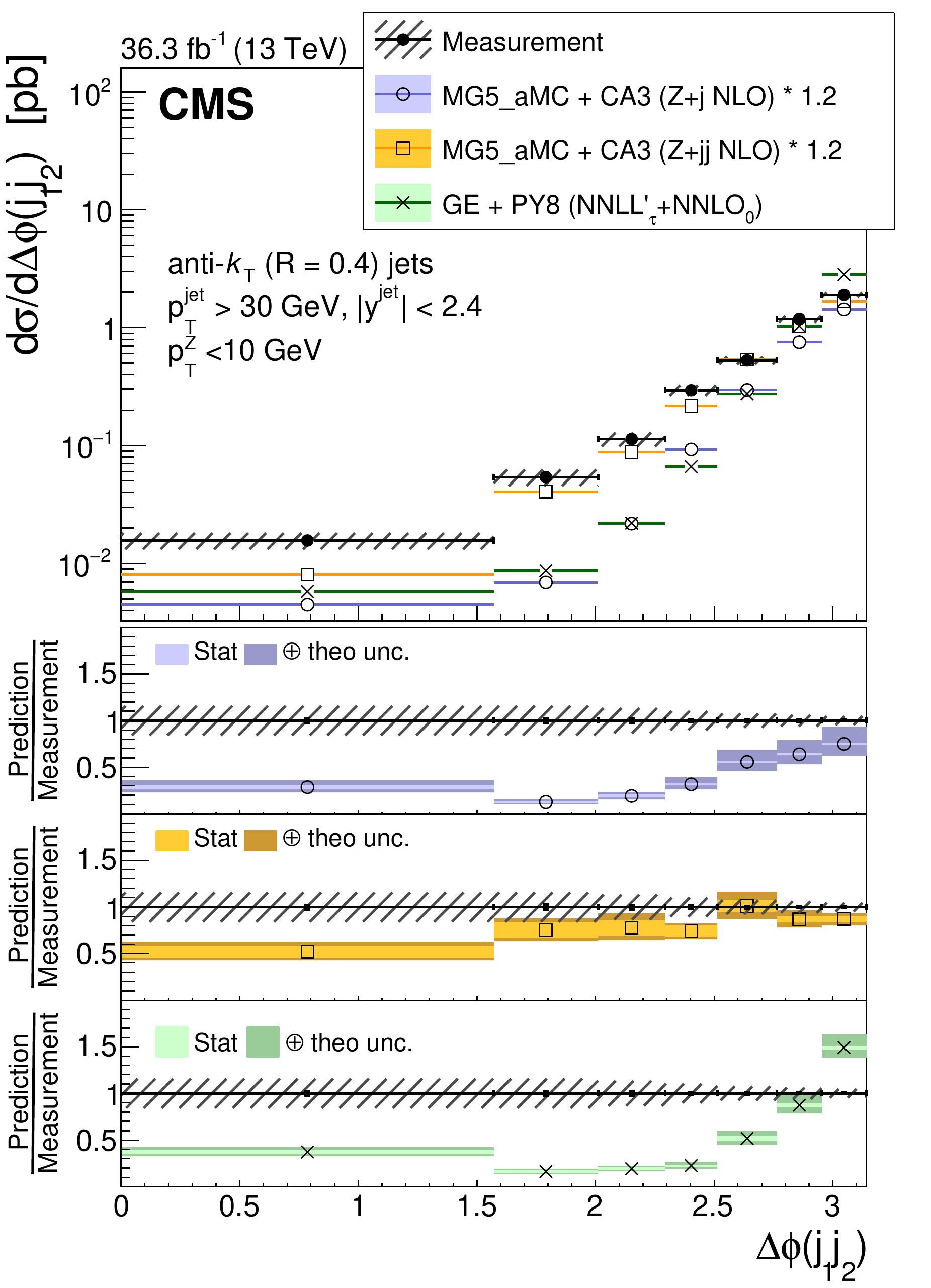}
    \includegraphics[width=0.488\textwidth]{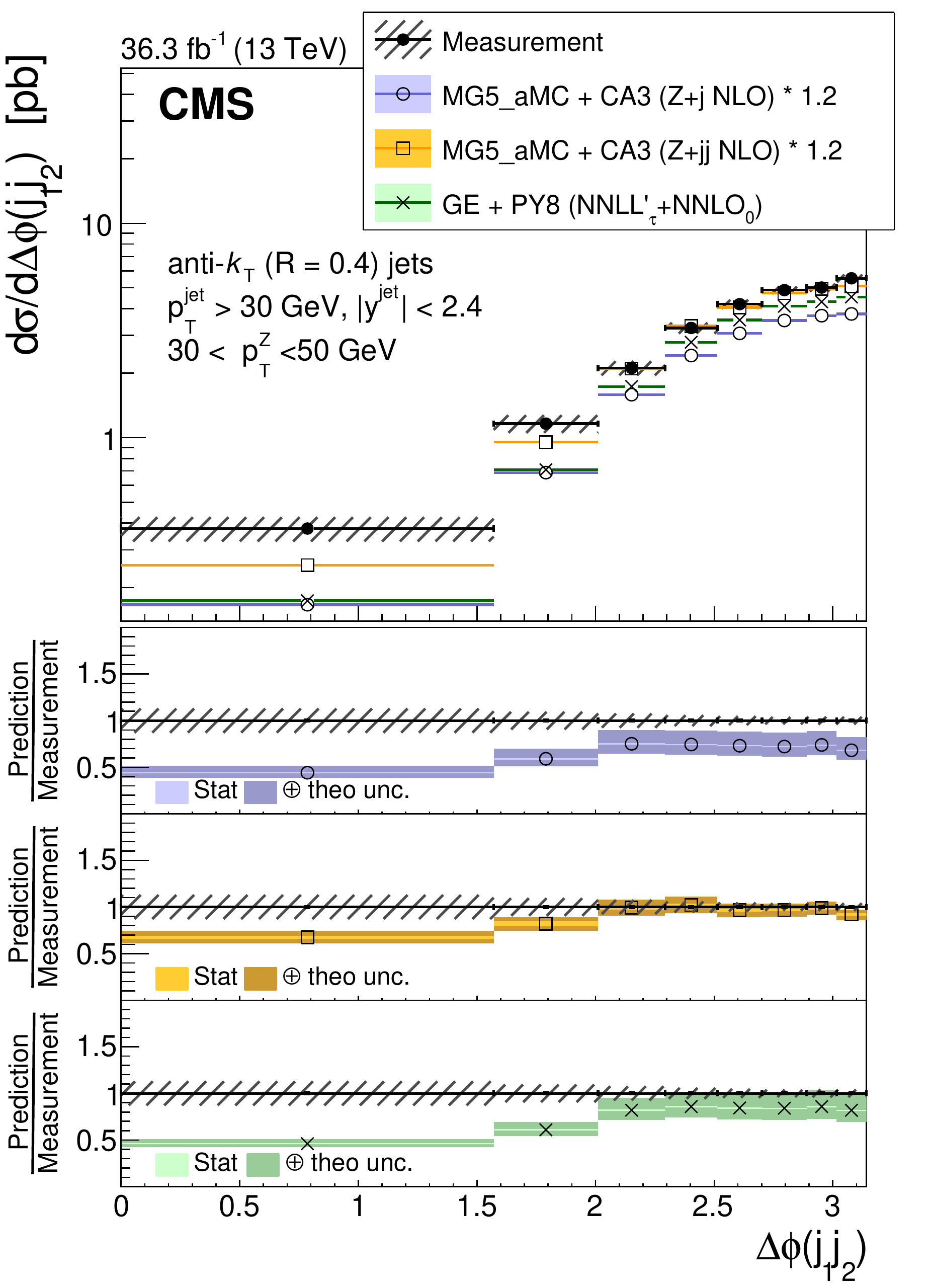}
    \includegraphics[width=0.488\textwidth]{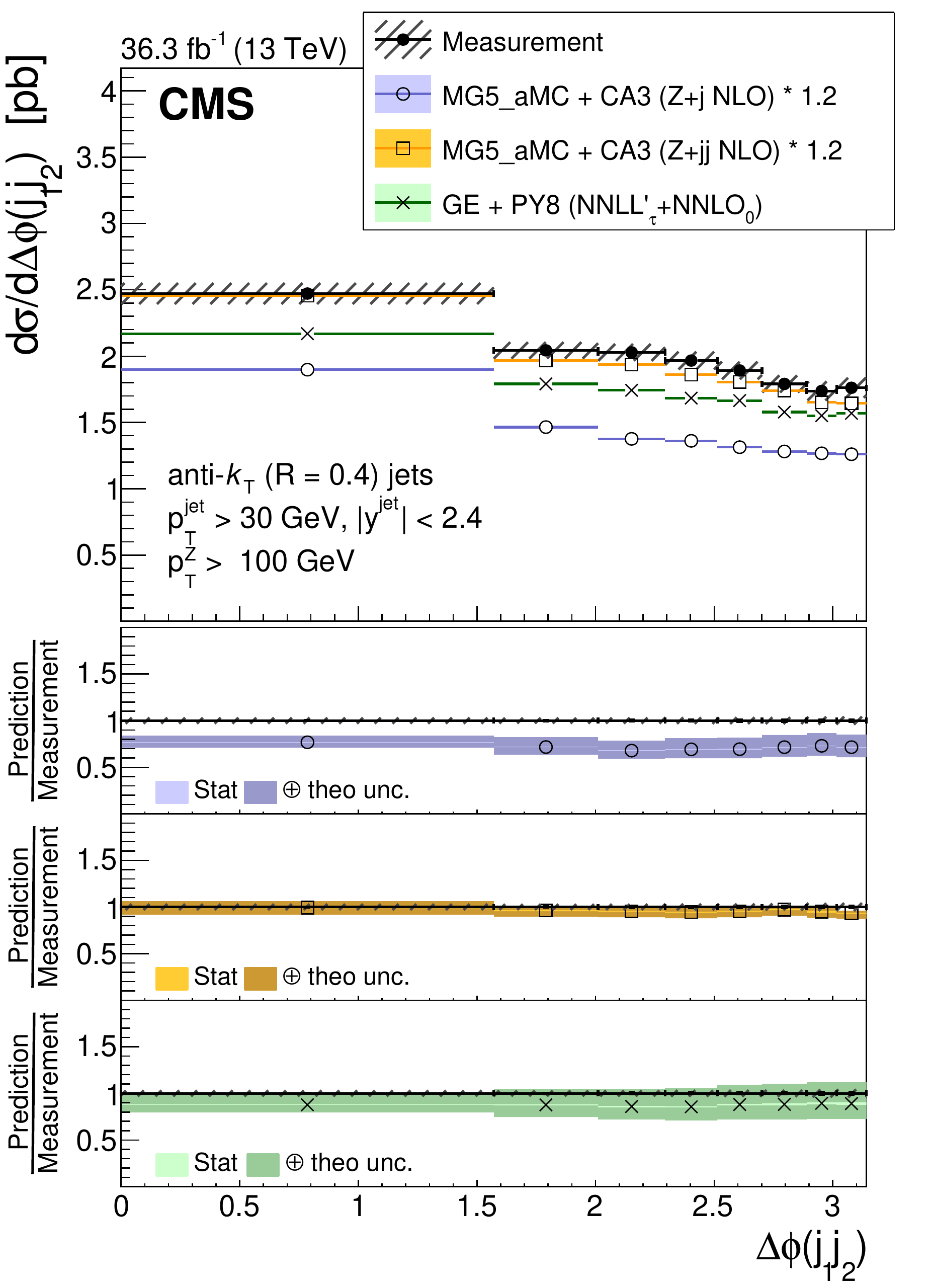}
    \topcaption{Cross section as a function of \dphiOneTwo between two leading jets in three \ptZ regions:
     $\ptZ <10\GeV$ (upper left), $30< \ptZ < 50\GeV$ (upper right),   $\ptZ > 100\GeV $ (lower).
  The error bars on the data points represent  the statistical uncertainty of the measurement, and the hatched band shows  the total statistical and systematic uncertainties added in quadrature.  
  Predictions from  \MGTMDZj,  \MGTMDZjj  and \geneva are shown. An overall normalization factor of 1.2 is applied to the \MGTMDZj  and \MGTMDZjj predictions.
   }
    \label{fig:muDPhiJJ-geneva-PB}
  \end{figure*}

{\tolerance=3000
In Fig.~\ref{fig:muDPhiJJ-geneva-PB}, the predictions from \MGTMDZj, \MGTMDZjj, and \geneva are shown. 
In general, the \MGTMDZj prediction is not sufficient to describe the measurement, whereas the \MGTMDZjj prediction describes the measurements at high \ptZ, where MPI effects are negligible. At lower \ptZ,  MPI effects become important, as shown in Fig.~\ref{fig:muDPhiJJ}.
 The \geneva prediction is below the measurement at low \ptZ because of missing higher order contributions, as is the prediction from \MGTMDZj.
\par}

\begin{figure*}[htbp]
\centering
  \includegraphics[width=0.488\textwidth]{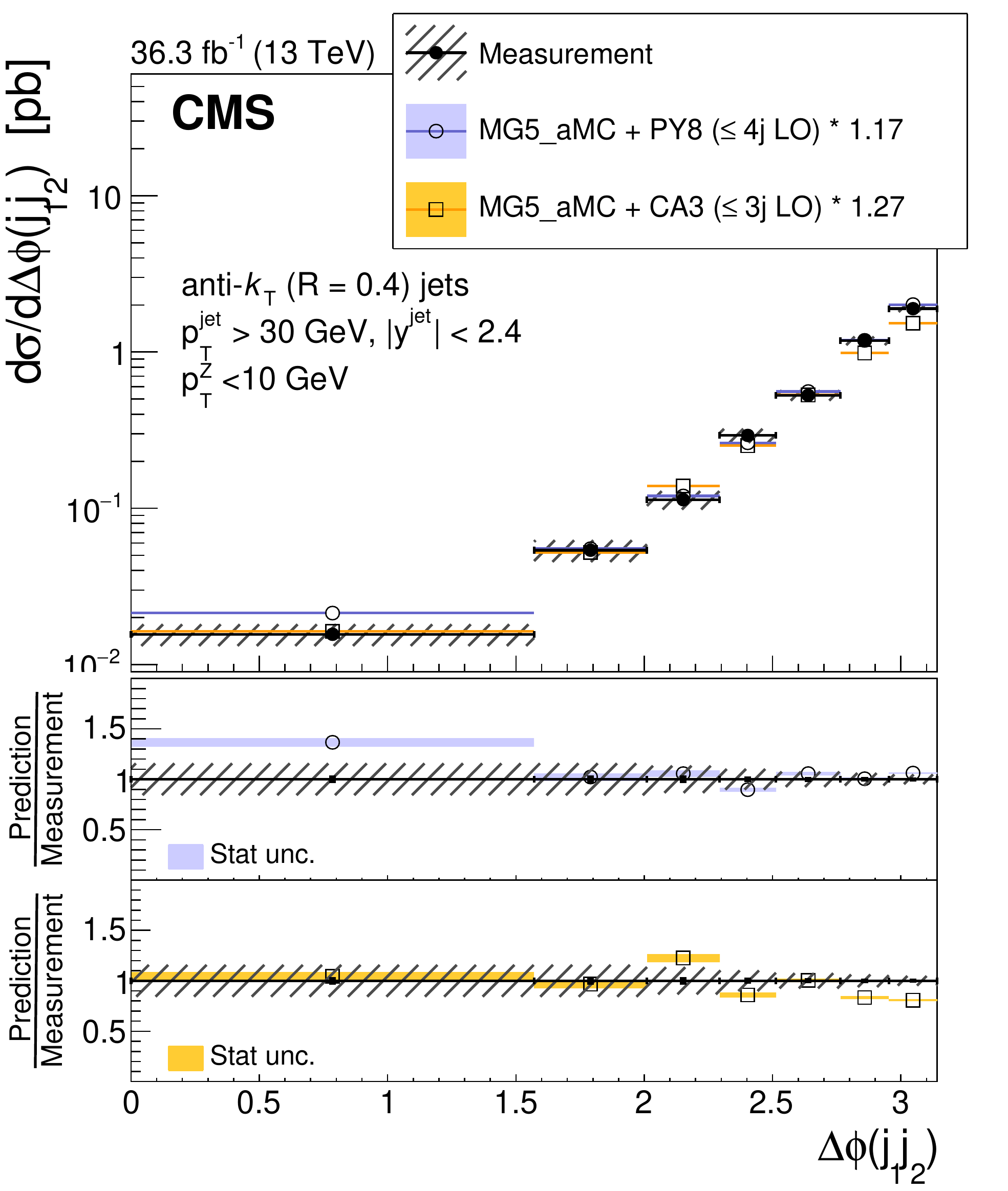}
  \includegraphics[width=0.488\textwidth]{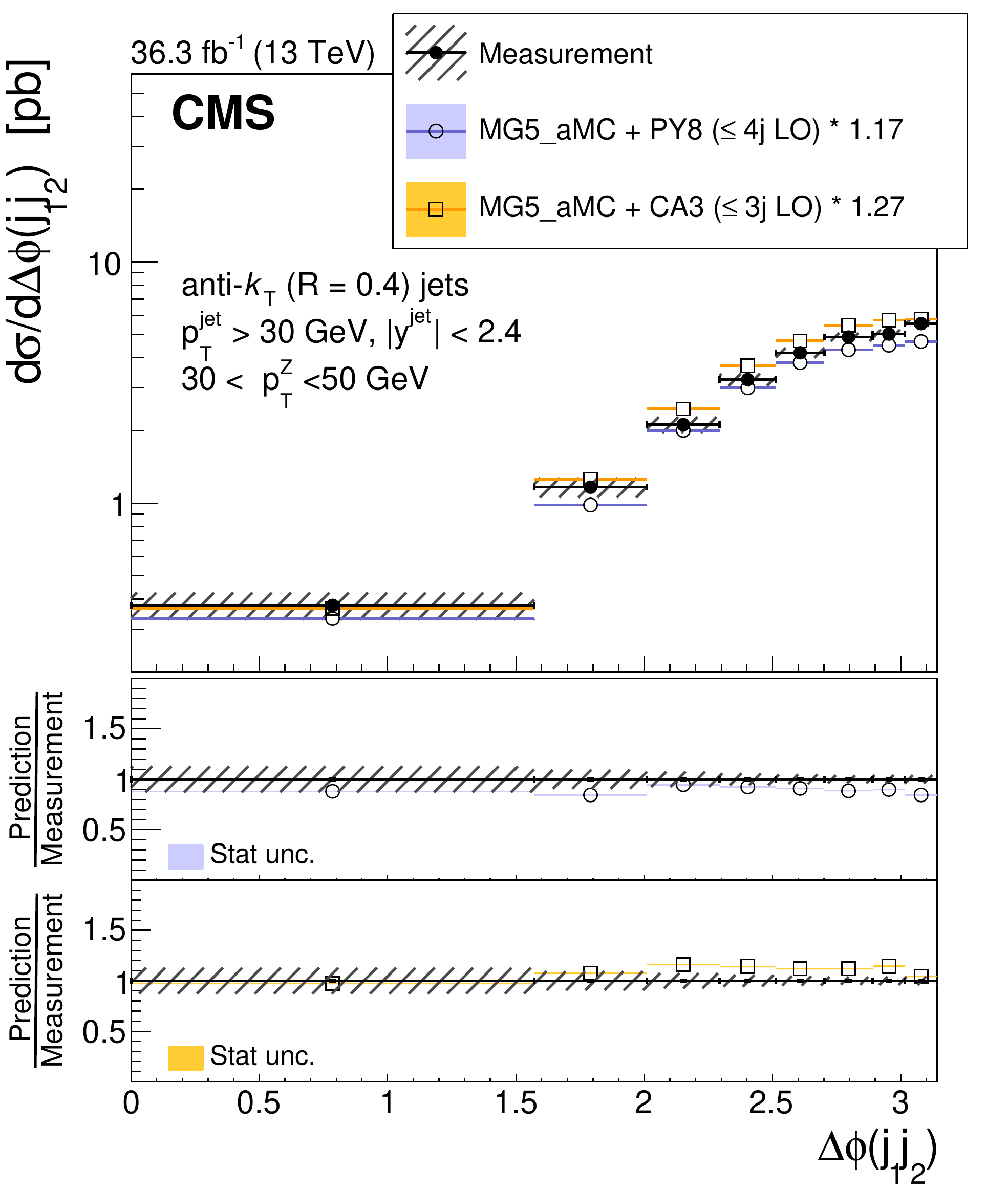}
  \includegraphics[width=0.488\textwidth]{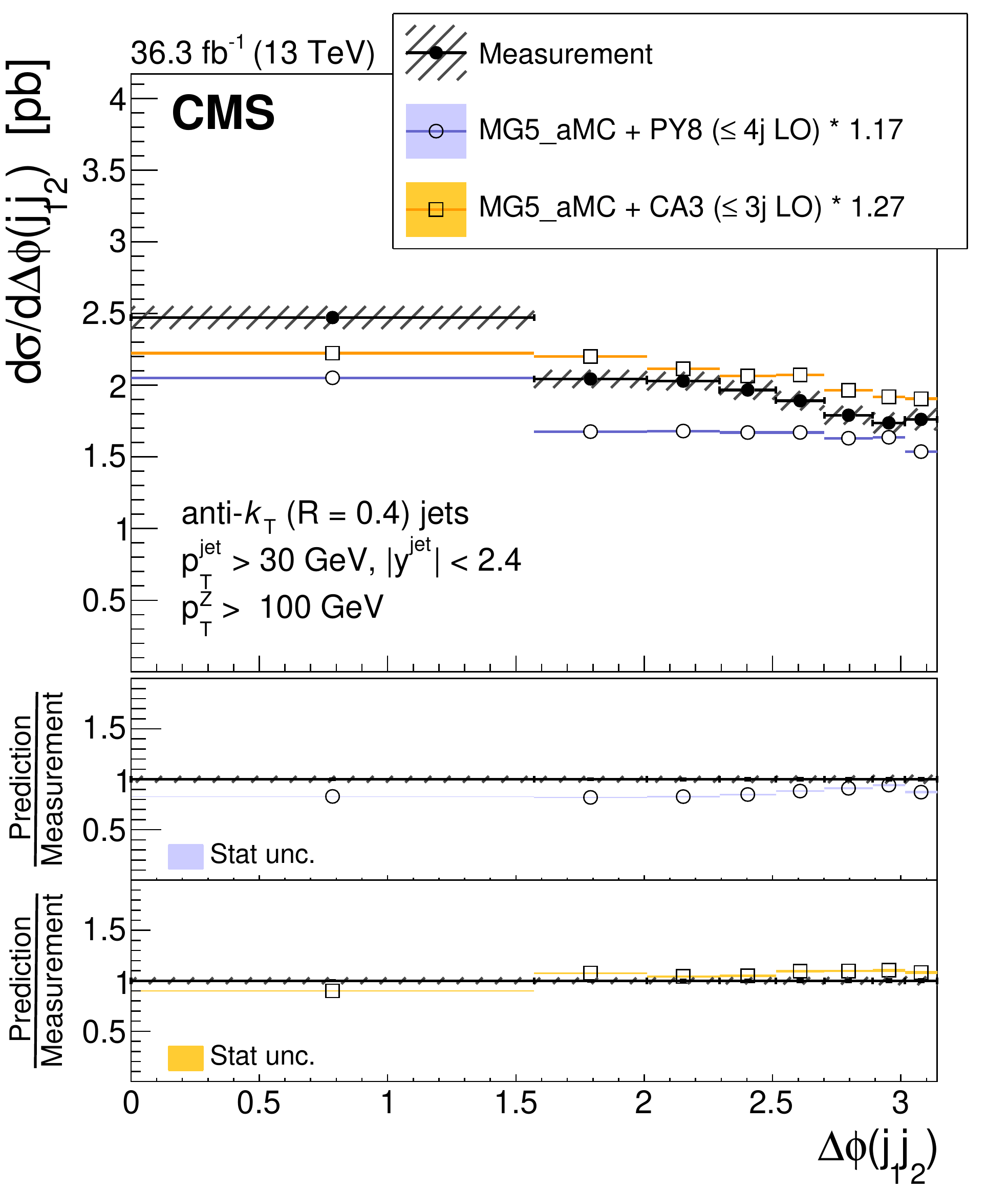}
  \topcaption{Cross section as a function of \dphiOneTwo between two leading jets in three \ptZ regions:
   $\ptZ <10\GeV$ (upper left), $30< \ptZ < 50\GeV$ (upper right),   $\ptZ > 100\GeV $ (lower).
The error bars on the data points represent  the statistical uncertainty of the measurement, and the hatched band shows  the total statistical and systematic uncertainties added in quadrature.  
Predictions from \MGPyLO  and \MGTMDMLM  are shown. Different normalization factors are applied, as described in the text.}
  \label{fig:muDPhiJJ-MLM}
\end{figure*}

In Fig.~\ref{fig:muDPhiJJ-MLM}, the predictions from \MGPyLO and \MGTMDMLM are compared with the measurement.
 The \MGPyLO prediction agrees with the measurement.
The \MGTMDMLM prediction describes the measurement in certain \ptZ regions.

  The contribution from MPI is significant  in the low \ptZ regions and becomes negligible when $\ptZ > 100\GeV$.
The calculation \MGPyNLO describes the measurements within the scale uncertainties, if an appropriate tune for PS and underlying event parameters are applied (here, the CUETP8M1 tune).

The predictions of  \MGTMD  using \PBM-TMD PDFs and initial-state \PBM PS come remarkably close to the measurements in regions of phase space where they are applicable.
 The prediction of \geneva for \PZ{}+jet observables describe the measurements in the regions of low jet multiplicity but show differences when two or more jets are selected since higher order contributions are not included.
The prediction of the merged  LO calculations \MGPyLO  and \MGTMDMLM describe the measurements quite well keeping in mind the role of MPI at low \ptZ.

{\tolerance=1000
In Ref.~\cite{Chien:2022wiq}, calculations using TMDs and the ``winner-takes-all" jet recombination scheme for the azimuthal angular decorrelation in \PZ{}+jet are described. The calculations reported here do not change significantly if the ``winner-takes-all" recombination scheme in the anti-\kt algorithm is applied. 
\par}

In this paper the differential cross section measurements are presented in three representative regions of the Z boson transverse momentum, 
the results for intervals $10<\ptZ<30\GeV$ and $50<\ptZ<100\GeV$ can be also found in HEPData~\cite{hepdata}.

\section{Summary} \label{sec:conclusions}

We have measured the  \PZ{}+jet production cross section in proton-proton collisions at the LHC at a center-of-mass energy of 13\TeV.
The associated jet multiplicity for various regions of the transverse momentum of the \PZ boson, \ptZ, was measured.
At $\ptZ < 10\GeV$ only about 1\% of the events have jets with $\pt > 30\GeV$, with nonnegligible cross sections at high jet multiplicity. At $30 < \ptZ < 50\GeV$, most of the events have at least one jet, with a significant tail to higher jet multiplicities. 
The azimuthal angle \dphiZj between the \PZ boson and the leading jet, as well as the azimuthal angle  \dphiOneTwo  between the two leading jets, was measured for the three \ptZ regions. At low \ptZ, the \PZ boson is only loosely correlated with the jets, but the two leading jets are strongly correlated. At large \ptZ, the \PZ boson is highly correlated with the leading jet, but the two leading jets are only weakly correlated.

The measurement shows that at low \ptZ  the \PZ boson appears as an electroweak correction to high-\pt jet production, whereas at large \ptZ the dominant process is \PZ{}+jet production.

{\tolerance=3000
The next-to-leading order (NLO) prediction of \MGPyNLO with \PZ{}+0,1,2 partons, which is merged with the FxFx procedure and supplemented with parton showering (PS) and multiple parton interactions (MPI) from \PYTHIAE, agrees with the measurements.
\par}
 
{\tolerance=4000
The predictions of \MGTMDZj and \MGTMDZjj using the parton branching method with transverse-momentum dependent (\PBM-TMD) parton densities, which do not
include MPI effects, and the corresponding PS agree with the measurements in the regions where MPI effects are negligible. The prediction from \geneva using matrix elements at next-to-next-to-leading order for \PZ production, supplemented with resummation, PS and MPI from \PYTHIAE, agrees with the measurements in the low jet multiplicity region.
\par}
 
{\tolerance=3000
The leading order prediction of \MGPyLO, including merging of jet multiplicities, describes the measurements well. 
 The prediction of \MGTMDMLM using \PBM-TMD parton densities and PS with merging of jet multiplicities agrees well with the measurements in the regions where MPI is negligible.
\par}

In summary, \PZ{}+jet measurements challenge theoretical predictions;  a good agreement can be achieved by including contributions of multiparton interactions, parton showering, parton densities, as well as multijet matrix element merging. The differential measurements provided here help to disentangle the various contributions and illustrate where each contribution becomes important. 
 
\begin{acknowledgments}
We congratulate our colleagues in the CERN accelerator departments for the excellent performance of the LHC and thank the technical and administrative staffs at CERN and at other CMS institutes for their contributions to the success of the CMS effort. In addition, we gratefully acknowledge the computing centers and personnel of the Worldwide LHC Computing Grid and other centers for delivering so effectively the computing infrastructure essential to our analyses. Finally, we acknowledge the enduring support for the construction and operation of the LHC, the CMS detector, and the supporting computing infrastructure provided by the following funding agencies: BMBWF and FWF (Austria); FNRS and FWO (Belgium); CNPq, CAPES, FAPERJ, FAPERGS, and FAPESP (Brazil); MES and BNSF (Bulgaria); CERN; CAS, MoST, and NSFC (China); MINCIENCIAS (Colombia); MSES and CSF (Croatia); RIF (Cyprus); SENESCYT (Ecuador); MoER, ERC PUT and ERDF (Estonia); Academy of Finland, MEC, and HIP (Finland); CEA and CNRS/IN2P3 (France); BMBF, DFG, and HGF (Germany); GSRI (Greece); NKFIH (Hungary); DAE and DST (India); IPM (Iran); SFI (Ireland); INFN (Italy); MSIP and NRF (Republic of Korea); MES (Latvia); LAS (Lithuania); MOE and UM (Malaysia); BUAP, CINVESTAV, CONACYT, LNS, SEP, and UASLP-FAI (Mexico); MOS (Montenegro); MBIE (New Zealand); PAEC (Pakistan); MES and NSC (Poland); FCT (Portugal); MESTD (Serbia); MCIN/AEI and PCTI (Spain); MOSTR (Sri Lanka); Swiss Funding Agencies (Switzerland); MST (Taipei); MHESI and NSTDA (Thailand); TUBITAK and TENMAK (Turkey); NASU (Ukraine); STFC (United Kingdom); DOE and NSF (USA).
  
\hyphenation{Rachada-pisek} Individuals have received support from the Marie-Curie program and the European Research Council and Horizon 2020 Grant, contract Nos.\ 675440, 724704, 752730, 758316, 765710, 824093, 884104, and COST Action CA16108 (European Union); the Leventis Foundation; the Alfred P.\ Sloan Foundation; the Alexander von Humboldt Foundation; the Belgian Federal Science Policy Office; the Fonds pour la Formation \`a la Recherche dans l'Industrie et dans l'Agriculture (FRIA-Belgium); the Agentschap voor Innovatie door Wetenschap en Technologie (IWT-Belgium); the F.R.S.-FNRS and FWO (Belgium) under the ``Excellence of Science -- EOS" -- be.h project n.\ 30820817; the Beijing Municipal Science \& Technology Commission, No. Z191100007219010; the Ministry of Education, Youth and Sports (MEYS) of the Czech Republic; the Hellenic Foundation for Research and Innovation (HFRI), Project Number 2288 (Greece); the Deutsche Forschungsgemeinschaft (DFG), under Germany's Excellence Strategy -- EXC 2121 ``Quantum Universe" -- 390833306, and under project number 400140256 - GRK2497; the Hungarian Academy of Sciences, the New National Excellence Program - \'UNKP, the NKFIH research grants K 124845, K 124850, K 128713, K 128786, K 129058, K 131991, K 133046, K 138136, K 143460, K 143477, 2020-2.2.1-ED-2021-00181, and TKP2021-NKTA-64 (Hungary); the Council of Science and Industrial Research, India; the Latvian Council of Science; the Ministry of Education and Science, project no. 2022/WK/14, and the National Science Center, contracts Opus 2021/41/B/ST2/01369 and 2021/43/B/ST2/01552 (Poland); the Funda\c{c}\~ao para a Ci\^encia e a Tecnologia, grant CEECIND/01334/2018 (Portugal); the National Priorities Research Program by Qatar National Research Fund; MCIN/AEI/10.13039/501100011033, ERDF ``a way of making Europe", and the Programa Estatal de Fomento de la Investigaci{\'o}n Cient{\'i}fica y T{\'e}cnica de Excelencia Mar\'{\i}a de Maeztu, grant MDM-2017-0765 and Programa Severo Ochoa del Principado de Asturias (Spain); the Chulalongkorn Academic into Its 2nd Century Project Advancement Project, and the National Science, Research and Innovation Fund via the Program Management Unit for Human Resources \& Institutional Development, Research and Innovation, grant B05F650021 (Thailand); the Kavli Foundation; the Nvidia Corporation; the SuperMicro Corporation; the Welch Foundation, contract C-1845; and the Weston Havens Foundation (USA).
\end{acknowledgments}

\bibliography{auto_generated}

\cleardoublepage \appendix\section{The CMS Collaboration \label{app:collab}}\begin{sloppypar}\hyphenpenalty=5000\widowpenalty=500\clubpenalty=5000
\cmsinstitute{Yerevan Physics Institute, Yerevan, Armenia}
{\tolerance=6000
A.~Tumasyan\cmsAuthorMark{1}\cmsorcid{0009-0000-0684-6742}
\par}
\cmsinstitute{Institut f\"{u}r Hochenergiephysik, Vienna, Austria}
{\tolerance=6000
W.~Adam\cmsorcid{0000-0001-9099-4341}, J.W.~Andrejkovic, T.~Bergauer\cmsorcid{0000-0002-5786-0293}, S.~Chatterjee\cmsorcid{0000-0003-2660-0349}, K.~Damanakis\cmsorcid{0000-0001-5389-2872}, M.~Dragicevic\cmsorcid{0000-0003-1967-6783}, A.~Escalante~Del~Valle\cmsorcid{0000-0002-9702-6359}, R.~Fr\"{u}hwirth\cmsAuthorMark{2}\cmsorcid{0000-0002-0054-3369}, M.~Jeitler\cmsAuthorMark{2}\cmsorcid{0000-0002-5141-9560}, N.~Krammer\cmsorcid{0000-0002-0548-0985}, L.~Lechner\cmsorcid{0000-0002-3065-1141}, D.~Liko\cmsorcid{0000-0002-3380-473X}, I.~Mikulec\cmsorcid{0000-0003-0385-2746}, P.~Paulitsch, F.M.~Pitters, J.~Schieck\cmsAuthorMark{2}\cmsorcid{0000-0002-1058-8093}, R.~Sch\"{o}fbeck\cmsorcid{0000-0002-2332-8784}, D.~Schwarz\cmsorcid{0000-0002-3821-7331}, S.~Templ\cmsorcid{0000-0003-3137-5692}, W.~Waltenberger\cmsorcid{0000-0002-6215-7228}, C.-E.~Wulz\cmsAuthorMark{2}\cmsorcid{0000-0001-9226-5812}
% ...
\par}
\vskip\cmsinstskip
\dag:~Deceased\\
$^{1}$Also at Yerevan State University, Yerevan, Armenia\\
$^{2}$Also at TU Wien, Vienna, Austria\\
% ...
\end{sloppypar}
\end{document}